\documentclass[10pt,sigconf]{acmart}

\usepackage{booktabs} 

\usepackage{times,amsmath,epsfig}
\usepackage[english]{babel}
\usepackage{graphicx}
\usepackage{tikz}
\usepackage{amssymb}
\usepackage{ifthen}
\usepackage{hyperref}
\usepackage{stmaryrd}
\usepackage[utf8x]{inputenc}
\usepackage{amsmath}
\usepackage{ifthen}
\usepackage{xcolor,colortbl}
\usepackage{graphicx}
\usepackage{balance}
\usepackage{marginnote}
\usepackage{multicol}
\usepackage{multirow}
\usepackage{listings}
\usepackage[linesnumbered]{algorithm2e}
\usepackage{algpseudocode}
\usepackage{mathtools}

\usepackage{bm}
\usepackage{caption} 
\usepackage{enumerate}
\usepackage{lipsum}
\usepackage[section]{placeins}
\usepackage{float}
\usepackage{verbatim}
\usepackage{setspace}
\usepackage{lipsum,graphicx,subcaption}

\newcommand{\calL}{\ensuremath{\mathcal L}}

\newcommand{\calD}{\ensuremath{\mathcal D}}

\newcommand{\calV}{\ensuremath{\mathcal V}}
\newcommand{\calE}{\ensuremath{\mathcal E}}

\newcommand{\darpe}{DARPE}
\newcommand{\daAlpha}{\ensuremath{\Sigma^{\{>,<\}}}}
\newcommand{\val}{\mbox{\tt val}}
\newcommand{\src}{\mbox{\tt src}}
\newcommand{\tgt}{\mbox{\tt tgt}}
\newcommand{\lab}{\ensuremath{\lambda}}
\newcommand{\select}{\mbox{SELECT}}
\newcommand{\AS}{\mbox{AS}}
\newcommand{\distinct}{\mbox{DISTINCT}}
\newcommand{\from}{\mbox{FROM}}
\newcommand{\where}{\mbox{WHERE}}
\newcommand{\accum}{\mbox{ACCUM}}
\newcommand{\paccum}{\mbox{POST\_ACCUM}}
\newcommand{\astmt}{acc-statement}
\newcommand{\groupby}{\mbox{GROUP\ BY}}
\newcommand{\having}{\mbox{HAVING}}
\newcommand{\orderby}{\mbox{ORDER\ BY}}
\newcommand{\limit}{\mbox{LIMIT}}

\newcommand{\length}[1]{|{#1}|}

\newcommand{\ctx}{\ensuremath{\mathit{ctx}}}
\newcommand{\dom}{\ensuremath{\mathit{dom}}}
\newcommand{\override}[2]{{#2} \triangleright {#1}}
\newcommand{\vIds}{\calV}
\newcommand{\eIds}{\calE}
\newcommand{\attrNames}{\ensuremath{\mathcal A}}
\newcommand{\vTyNames}{\ensuremath{\mathcal T}_v}
\newcommand{\eTyNames}{\ensuremath{\mathcal T}_e}

\newcommand{\epts}{\ensuremath{\mathit{st}}}
\newcommand{\tyV}{\ensuremath{\tau_v}}
\newcommand{\tyE}{\ensuremath{\tau_e}}
\newcommand{\ty}[1]{\ensuremath{{#1}.\mathit{type}}}
\newcommand{\data}{\ensuremath{\delta}}
\newcommand{\gAccNames}{\ensuremath{\mathit{Acc}_g}}
\newcommand{\vAccNames}{\ensuremath{\mathit{Acc}_v}}

\newcommand{\mapGAcc}{\ensuremath{\mu_{\mathit{gacc}}}}
\newcommand{\mapVAcc}{\ensuremath{\mu_{\mathit{vacc}}}}
\newcommand{\deltaG}{\ensuremath{\eta_{\mathit{gacc}}}}
\newcommand{\deltaV}{\ensuremath{\eta_{\mathit{vacc}}}}

\newcommand{\bag}[1]{\ensuremath{\{\!|{#1}|\!\}}}
\newcommand{\sem}[2]{\ensuremath{[\![{#2}]\!]^{#1}}}
\newcommand{\binding}[2]{\ensuremath{\beta_{#1}^{#2}}}
\newcommand{\incr}{\ensuremath{+\!\!=}}

\newcommand{\reduce}[1]{\ensuremath{\mathit{reduce}_{#1}}}

\newcommand{\mybf}[1]{\ensuremath{\pmb{\bm{#1}}}}

\usepackage{tikz}
\usepackage{lipsum,graphicx,multicol,multirow}

\usetikzlibrary{arrows, automata}
\usetikzlibrary{positioning,shapes,shadows,arrows,decorations.markings}
\usetikzlibrary{decorations.pathmorphing}
\usepackage{tikz-uml}
\newcommand{\reminder}[1]{[\vadjust{\vbox to0pt{\vss\hbox to0pt{\hss{\Large $\Longrightarrow$}}}}{{\textsf{\small #1}}}]}
\newcommand{\AD}[1]{\textcolor{green}{\reminder{alin:~#1}}}
\renewcommand{\AD}[1]{}

\newcommand{\eat}[1]{}
\newcommand{\myparagraph}[1]{\vspace{1.5 mm}\noindent\textbf{{#1}.~}}
\newcommand{\myhref}[2]{\href{#1}{\color{blue}{#2}}}

\newtheorem{Example}{\textbf{Example}}

\newenvironment{myexample}[1]{\noindent\begin{Example}[{\bf {#1}}]}{\hfill $\Box$\end{Example}}

\setcopyright{rightsretained}

\newcommand{\id}{{\tt id}}

\setcopyright{none}

\acmArticle{4}
\acmPrice{15.00}

\begin{document}

\settopmatter{printacmref=false}
\title {TigerGraph: A Native MPP Graph Database}
       
\author{Alin Deutsch \hspace{13mm}Yu Xu\hspace{13mm}Mingxi Wu\hspace{13mm}Victor Lee}
\affiliation{%
  \institution{\hspace{3mm}UC San Diego\hspace{13mm} TigerGraph\hspace{13mm} TigerGraph\hspace{17mm} TigerGraph}
}
\email{deutsch@cs.ucsd.edu, {yu, mingxi.wu, victor}@tigergraph.com}

\renewcommand{\shortauthors}{Deutsch, Xu, Wu and Lee}

\begin{abstract}
  We present TigerGraph, a graph database system built from the ground up to
  support massively parallel computation of queries and analytics.
  TigerGraph's high-level query language, GSQL, is designed for
  compatibility with SQL, while simultaneously allowing NoSQL programmers
  to continue thinking in Bulk-Synchronous Processing (BSP) terms and reap the
  benefits of high-level specification.
  GSQL is sufficiently high-level to allow declarative SQL-style programming, yet sufficiently expressive
  to concisely specify the sophisticated iterative algorithms required by modern graph analytics and
  traditionally coded in general-purpose programming languages like C++ and Java.
  We report very strong scale-up and scale-out performance  over a benchmark we
  published on GitHub for full reproducibility.
\end{abstract}

%
%
\eat{
}


\lstdefinestyle{GSQL}{
  language=SQL,
  otherkeywords={QUERY, FOR, GRAPH, WITH, BEGIN, END, ACCUM, POST, WHILE, DO, VERTEX, vertex, RETURN, TO, UNDIRECTED, DIRECTED, UN, EDGE, REVERSE, STRING, SumAccum, MaxAccum, log, abs, += },
  numbers=left,
  stepnumber=0,
  numbersep=10pt,
  tabsize=4,
  showspaces=false,
  showstringspaces=false
}
\lstset{
    escapeinside={(*@}{@*)},
}

\lstdefinestyle{Grammar}{
  language=SQL,
  otherkeywords={HAVING, QUERY, FOR, for, GRAPH, WITH, BEGIN, END, ACCUM, POST, WHILE, while, break, continue, DO, IF, if,  VERTEX, vertex, RETURN, TO, UNDIRECTED, DIRECTED, UN, EDGE, REVERSE, DISCRIMINATOR, STRING, BETWEEN, IN, LIKE, IS, is, INTERSECT, intersect, TERSECT, tersect, DO, do, SetAccum, BagAccum, HeapAccum, OrAccum, AndAccum, MaxAccum, MinAccum, SumAccum, AvgAccum, ListAccum, ArrayAccum, StringAccum, MapAccum, GroupByAccum, ASC, DESC, +=, =, -, ., :, ;, _, <, >, @, ', Id, NumConst, StringConst, DateTimeConst, true, false, type, int, long, float, double, string, boolean, datetime, Tuple, set, bag, map},
  numbers=left,
  stepnumber=0,
  numbersep=10pt,
  tabsize=4,
  showspaces=false,
  showstringspaces=false
}

\lstset{
    escapeinside={(*@}{@*)},
}

\maketitle
\section{Introduction}

Graph database technology is among the fastest-growing segments in today's data management
industry.
Since seeing early adoption by companies including Twitter, Facebook and Google,
graph databases have evolved into a mainstream technology used today by enterprises
across industries, complementing (and sometimes replacing)
both traditional RDBMSs and newer NoSQL big-data products.
Maturing beyond social networks,
the technology is disrupting an increasing number of areas, such as supply chain management,
e-commerce recommendations, cybersecurity, fraud detection, power grid monitoring,
and many other areas in advanced data analytics.

While research on the graph data model (with associated query languages and academic
prototypes) dates back to the late 1980s, in recent years we have witnessed the
rise of several products offered by commercial software companies like Neo Technologies
(supporting the graph query language Cypher~\cite{neo4j}) and DataStax~\cite{datastax}
(supporting Gremlin~\cite{gremlin}).
These languages
are also supported by the commercial offerings of many other companies
(Amazon Neptune~\cite{neptune}, IBM Compose for JanusGraph~\cite{ibmcompose},
Microsoft Azure CosmosDB~\cite{cosmosdb}, etc.).

We introduce TigerGraph, a new graph database product by the homonymous company.
TigerGraph is a {\em native parallel graph database}, in the sense that its proprietary
storage is designed from the ground up to store graph nodes, edges and their attributes
in a way that supports an engine that computes queries and analytics in
massively parallel processing (MPP) fashion for significant
scale-up and scale-out performance.

TigerGraph allows developers to express
queries and sophisticated graph analytics using a high-level language called GSQL.
We subscribe to the requirements for a modern query language listed in the
G-Core manifesto~\cite{gcore}. To these, we add
\begin{itemize}
\item
  facilitating adoption by the largest query developer community in existence, namely SQL
  developers. GSQL was designed for full compatibility with SQL in the sense that if
  no graph-specific primitives are mentioned, the queries become pure standard SQL. The
  graph-specifc primitives include the flexible regular path expression-based patterns
  advocated in~\cite{gcore}. 
\item
  the support, beyond querying, of classical multi-pass and iterative algorithms as
  required by modern graph analytics (such as PageRank, weakly-connected components,
  shortest-paths, recommender systems, etc., all GSQL-expressible). This is achieved
  while staying declarative and high-level by introducing only two primitives: loops and
  accumulators.
\item
  allowing developers with NoSQL background (which typically espouses a
  low-level and imperative programming style) to preserve their Map/Reduce or graph
  Bulk-Synchronous Parallel (SP)~\cite{bsp} mentality while reaping the benefits of
  high-level declarative query specification.
  GSQL admits natural Map/Reduce and graph BSP interpretations and is actually implemented
  accordingly to support parallel processing.
\end{itemize}

A free TigerGraph developer edition can be
downloaded from the company Web site~\footnote{\myhref{http://tigergraph.com}{http://tigergraph.com}},
together with documentation, an e-book, white papers,
a series of representative analytics examples reflecting real-life customer needs,
as well as the results of a benchmark comparing our engine to other commercial
graph products.

\eat{
The FROM, and WHERE clauses are used to select and filter a set of edges or nodes.
After this selection, the optional ACCUM clause defines a set of
actions to be performed by each edge or adjacent node.
We say “perform by” rather than “perform on” because in this interpretation, each graph object
(node or edge) is an independent computation unit. 
An ACCUM clause actions can read values from
the graph objects, perform local computations, apply conditional statements, and schedule updates of the
graph. To support these distributed computations, the GSQL language provides accumulator
variables, which aggregate the values generated by the parallel computing units.
For example, a simple sum accumulator would be used to perform the count of all the neighbors’ neighbors
mentioned above. A set accumulator would be used to record the IDs of all those neighbors’ neighbors.
Accumulators are available in two scopes: global and per-node.
In the earlier query example, we mentioned the
need to mark each node as visited or not. Here, per-node accumulators would be used.
}

\myparagraph{Paper Organization}
The remainder of the paper is organized as follows.
Section~\ref{sec:design-overview} overviews TigerGraph's key design choices, while
its architecture is described in Section~\ref{sec:architecture}. We present the DDL and
DML in Sections~\ref{sec:ddl} and \ref{sec:dml}, respectively. We discuss evaluation complexity
in Section~\ref{sec:complexity},
BSP interpretations in Section~\ref{sec:mpp-interpretation},
and we conclude in Section~\ref{sec:conclusions}. 
In Appendix~\ref{sec:benchmark}, we report on an experimental evaluation using
a benchmark published for 
reproducibility in TigerGraph's
\myhref{https://github.com/tigergraph/ecosys/tree/benchmark/benchmark/tigergraph}{GitHub}
repository.

\section{Overview of TigerGraph's Native Parallel Graph Design}
\label{sec:design-overview}

We overview here the main ingredients of TigerGraph's design,
showing how they work together to achieve speed and scalability.

\myparagraph{A Native Distributed Graph}
TigerGraph was designed from the ground up as a native graph database.
Its proprietary data store holds nodes, edges, and their attributes.
We decided to avoid the more facile solution of building a wrapper on top of a more generic NoSQL
data store because this virtual graph strategy incurs a double performance penalty.
First, the translation from virtual graph manipulations to physical storage operations introduces
overhead. Second, the underlying structure is not optimized for graph operations. 

\myparagraph{Compact Storage with Fast Access}
TigerGraph is not an in-memory database, because holding all data in memory is a preference but not
a requirement (this is true for the enterprise edition, but not the free developer edition).
Users can set parameters that specify how much of the available memory may be used for holding
the graph. If the full graph does not fit in memory, then the excess is spilled to disk.

Data values are stored in encoded formats that effectively compress the data.
The compression factor varies with the graph structure and data, but typical
compression factors are between 2x and 10x. Compression
reduces not only the memory footprint and thus the cost to users, but also CPU cache misses,
speeding up overall query performance.  
Decompression/decoding is efficient and transparent to end users,
so the benefits of compression outweigh the
small time delay for compression/decompression. In general, the encoding is
{\em homomorphic}~\cite{homomorphic-encoding}, 
that is decompression is needed only for displaying the
data. When values are used internally, often they may remain encoded and compressed.
Internally hash indices are used to reference nodes and edges.
For this reason, accessing a particular node or edge in the graph is fast,
and stays fast even as the graph grows in size. Moreover, maintaining the index as
new nodes and edges are added to the graph is also very fast.

\myparagraph{Parallelism and Shared Values}
TigerGraph was built for parallel execution, employing a design that supports
massively parallel processing (MPP) in
all aspects of its architecture. 
TigerGraph exploits the fact that graph queries are compatible with parallel computation.
Indeed, the nature of graph queries is to follow the edges between nodes, traversing multiple distinct
paths in the graph. These traversals are a natural fit for parallel/multithreaded execution.
Various graph algorithms require these traversals to proceed according to certain disciplines,
for instance in a breadth-first manner,
keeping book of visited nodes and pruning traversals upon encountering them.
The standard solution is to assign a temporary variable to each node, marking whether
it has already been visited.
While such  marker-manipulating operations may suggest low-level,
general-purpose programming languages, one can actually express
complex graph traversals in a few lines of code
(shorter than this paragraph) using TigerGraph’s high-level query language.

\myparagraph{Storage and Processing Engines Written in C++}
TigerGraph’s storage engine and processing engine are
implemented in C++. A key reason for this choice is the fine-grained control of memory
management offered by C++.
Careful memory management contributes to TigerGraph’s ability to traverse many edges simultaneously
in a single query. While an alternative implementation using an
already provided memory management (like in the Java Virtual Machine) would be convenient,
it would make it difficult for the programmer to optimize memory usage.

\myparagraph{Automatic Partitioning}
In today’s big data world, enterprises need their database solutions to be able to scale out
to multiple machines, because their data may grow too large to be stored economically on a single
server. TigerGraph is designed to automatically partition the graph data across a cluster of servers
to preserve high performance. The hash index is used to determine not only the within-server  but also
the which-server data location. All the edges that connect out from a given node are stored on the
same server.

\myparagraph{Distributed Computing}
TigerGraph supports a distributed computation mode that significantly improves performance for
analytical queries that traverse a large portion of the graph. In distributed query mode, all
servers are asked to work on the query; each server’s actual participation is on an as-needed
basis. When a traversal path crosses from server A to server B, the minimal amount of information
that server B needs to know is passed to it. Since server B already knows about the overall query
request, it can easily fit in its contribution. In a benchmark study, we tested the commonly used
PageRank algorithm. This algorithm is a severe test of a graph’s computational and
communication speed because it traverses every edge, computes a score for every
node, and repeats this traverse-and-compute for several iterations.
When the graph was distributed across eight servers compared to a single-server, the PageRank
query completed nearly seven times faster (details in Appendix~\ref{sec:benchmark-scale-out}).
This attests to TigerGraph's efficient use of distributed infrastructure.

\myparagraph{Programming Abstraction: an MPP Computation Model}\\
The low-level programming abstraction offered by TigerGraph
integrates and extends the two classical graph programming paradigms of
think-like-a-vertex (a la Pregel~\cite{pregel}, GraphLab~\cite{graphlab} and Giraph~\cite{giraph})
and think-like-an-edge (PowerGraph~\cite{powergraph}).
Conceptually, each node or edge acts simultaneously as a parallel unit of storage and computation,
being associated with a compute function programmed by the user.
The graph becomes a massively parallel computational mesh that
implements the query/analytics engine.

\myparagraph{GSQL, a High-Level  Graph Query Language}
TigerGraph offers its own high-level
graph querying and update language, GSQL. 
The core of most GSQL queries is the SELECT-FROM-WHERE block,
modeled closely after SQL. GSQL queries also feature two
key extensions that support efficient parallel computation:
a novel \accum\ clause that specifies node and edge compute functions,
and accumulator variables
that aggregate inputs produced by parallel executions of these compute functions.
GSQL queries have a declarative semantics that
abstracts from the underlying infrastructure and is compatible with SQL.
In addition, GSQL queries admit equivalent MPP-aware interpretations that appeal to NoSQL
developers and are exploited in implementation.

\section{System Architecture}
\label{sec:architecture}

\begin{figure}[t]
    \includegraphics[width=1\columnwidth,trim={7 40 10 40},clip=true]{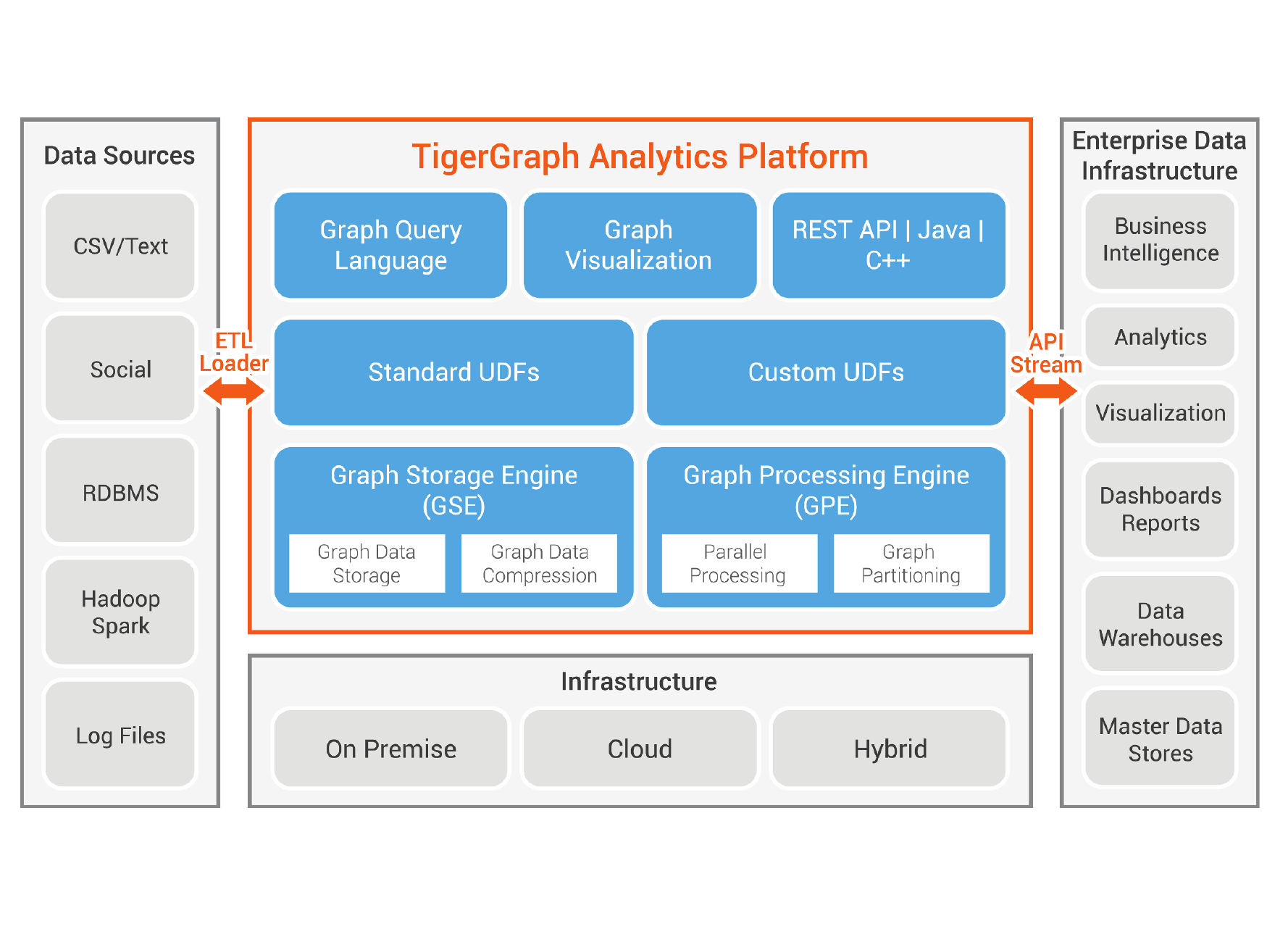}
    \caption{TigerGraph System Architecture}
    \label{fig:architecture}
\end{figure}

TigerGraph's architecture is depicted by Figure \ref{fig:architecture} (in the blue boxes).
The system vertical stack is structured in three layers:
the top layer comprises the user programming interface,
which includes the GSQL compiler, the GraphStudio visual interface, and REST APIs
for other languages to connect to TigerGraph; the middle layer contains
the standard built-in and user defined functions (UDFs); the bottom
layer includes the graph storage engine (GSE) and the graph processing
engine (GPE). We elaborate on each component next.

\vspace{1ex}\noindent\textbf{The GSQL compiler} is responsible
for query plan generation. Downstream it can send the query plan to
a query interpreter, or it can generate native machine code based on the query plan.
The compiler modules perform type checks, semantic checks, query transformations,
plan generation, code-generation and catalog management.
The GSQL compiler supports syntax for the Data Definition Language (DDL),
Data Loading Language (DLL), and Data Manipulation Language (DML).
A GSQL query text can be input either by typing into the GSQL shell console or
via a REST API call. Once a query plan is generated by the compiler, 
depending on the user's choice it can be sent
to an interpreter to evaluate the query in real-time. Alternatively,
the plan can be sent to the code generation module, where C++ UDFs and a REST endpoint
are generated and compiled into a dynamic link library. If the compiled mode is chosen,
the GSQL query is automatically deployed as a service that is invokable with different
parameters via the REST API or the GSQL shell. 

\vspace{1ex}\noindent\textbf{The GraphStudio } visual SDK is a simple yet powerful graphical
user interface. GraphStudio integrates all the phases of graph data
analytics into one easy-to-use graphical web-browser user interface.
GraphStudio is useful for ad hoc, interactive analytics and for learning
to use the TigerGraph platform via drag-and-drop query building. Its components include
a schema builder, a visual data loader, a graph explorer and the GSQL visual
IDE. These components (except for the graph explorer) talk directly
to the GSQL compiler. The graph explorer talks to the underlying engine via
TigerGraph system built-in standard UDFs. 

\vspace{1ex}\noindent\textbf{The REST APIs } are the interfaces for other third party processes
talking to TigerGraph system. The TigerGraph system includes a REST server
programmed and optimized for high-throughput REST calls. All user defined queries,
topology CRUD (create, read, update, delete) standard operations, and data ingestion loading jobs
are installed as REST endpoints. The REST server will process all
the http requests, going through request validation,
dispatching, and relaying JSON-formatted responses to the http client. 

\vspace{1ex}\noindent\textbf{The Standard and Custom UDFs } are C++ functions encoding
application logic. These UDFs serve as a bridge between the upper
layer and the bottom layer. UDFs can reside in static library form, being linked
at engine compile time. Alternatively, they can be a dynamic linked library hookable
at the engine runtime. Many
standard graph operations and generic graph algorithms are coded in pre-built
static libraries and linked with the engine at each system release.
Ad hoc user queries are generated at GSQL compile time and linked to
the engine at runtime.
\eat{
  Regardless which form, these UDFs follow a
  EdgeMap-VertexReduce scheme. An EdgeMap function specifies the user
  processing logic on a set of edges. The VertexReduce function describes
  the user processing logic on each vertex's receiving messages. A
  user query is essentially translated into a series of user-defined
  EdgeMap-VertexReduce functions, which are processed by the underlying
  parallel engine. 
}

\vspace{1ex}\noindent\textbf{GSE } is the storage engine. It is responsible for ID management,
meta data management, and topology storage management in different
layers (cache, memory and disk) of the hardware. ID management involves
allocating or de-allocating internal ids for graph elements (vertex and
edge). Meta data management stores catalog information, persists
system states and synchronizes different system components to collaboratively
finish a task. Topology storage management focuses on minimizing the
memory footprint of the graph topology description while maximizing the parallel
efficiency by slicing and compressing the graph into ideal
parallel compute and storage units. It also provides kernel APIs to answer
CRUD requests on the three data sets that we just mentioned. 

\vspace{1ex}\noindent\textbf{GPE }is the parallel engine. It processes
UDFs in a bulk synchronous parallel (BSP) fashion. GPE manages the
system resources automatically, including memory, cores, and machine
partitions if TigerGraph is deployed in a cluster environment (this is supported
in the Enterprise Edition). Besides resource management,
GPE is mainly responsible for parallel processing of tasks from the
task queue. A task could be a custom UDF, a standard UDF
or a CRUD operation. GPE synchronizes, schedules and processes all
these tasks to satisfy ACID transaction semantics while maximizing
query throughput. 
Details can be found online~\footnote{
\myhref{https://docs.tigergraph.com/dev/gsql-ref/querying/distributed-query-mode}{https://docs.tigergraph.com/dev/gsql-ref/querying/distributed-query-mode}}.

\myparagraph{Components Not Depicted}
There are other miscellaneous components that are not depicted in the architecture
diagram, such as a Kafka message service for buffering query requests
and data streams, a control console named GAdmin for the system admin
to invoke kernel function calls, backup and restore, etc.

\section{GSQL DDL}
\label{sec:ddl}

In addition to the type-free setting, GSQL also supports an SQL-like mode with a strong type system.
This may seem surprising given that prior work on graph query languages
traditionally touts schema freedom as a desirable feature
(such data was not accidentally called ``semi-structured'').
However, this feature was historically motivated by the driving application
at the time, namely integrating numerous heterogeneous, third-party-owned
but individually relatively small data sources from the Web.

In contrast, TigerGraph targets enterprise applications,
where the number and ownership of sources is not a concern,
but their sheer size and resulting performance challenges are.
In this setting, vertices and edges that model the same
domain-specific entity tend to be uniformly structured and advance
knowledge of their type is expected.
Failing to exploit it for performance would be a missed opportunity.

GSQL's Data Definition Language (DDL) shares SQL's philosophy of
defining in the same CREATE statement a persistent
data container as well as its type. This is in contrast to
typical programming languages which define types as stand-alone
entities, separate from their instantiations.

GSQL's CREATE statements can define
vertex containers, edge containers, and graphs consisting of these
containers.
The attributes for the vertices/edges
populating the container are declared using syntax borrowed from SQL's
CREATE TABLE command.
In TigerGraph's model, a graph may contain both directed and undirected
edges. The same vertex/edge container may be shared by multiple graphs.

\begin{myexample}{DDL}
  \label{ex:ddl}
  Consider the following DDL statements declaring two graphs:
  LinkedIn and Twitter. Notice how symmetric relationships (such
  as LinkedIn connections) are modeled as undirected edges, while
  asymmetric relationships (such as Twitter's following or posting)
  correspond to directed edges. Edge types specify the types of
  source and target vertices, as well as optional edge attributes
  (see the {\tt since} and {\tt end} attributes below).

  For vertices, one can declare primary key attributes with the
  same meaning as in SQL (see the {\tt email} attribute of Person vertices).

\lstset{basicstyle=\fontsize{8.5}{7.5}\ttfamily,style=GSQL}  
\begin{lstlisting}
CREATE VERTEX Person
   (email STRING PRIMARY KEY, name STRING, dob DATE)

CREATE VERTEX Tweet
   (id INT PRIMARY KEY, text STRING,(*@ timestamp @*) DATE)

CREATE DIRECTED EDGE Posts
   (FROM Person, TO Tweet)

CREATE DIRECTED EDGE Follows
   (FROM Person, TO Person, since DATE)

CREATE UNDIRECTED EDGE Connected
   (FROM Person, TO Person, since DATE,(*@ end @*)DATE)

CREATE GRAPH LinkedInGraph
   (Person, Connected)
CREATE GRAPH TwitterGraph
   (Person, Tweet, Posts, Follows)
\end{lstlisting}
\end{myexample}

By default, the primary key of an edge type is the composite key
comprising the primary keys of its endpoint vertex types.

\myparagraph{Edge Discriminators}
Multiple parallel edges of the same type between the
same endpoints are allowed. To distinguish among them, one can
declare {\em discriminator attributes} which complete the
pair of endpoint vertex keys to uniquely identify the edge.
This is analogous to the concept of weak entity
set discriminators in the Entity-Relationship Model~\cite{ER}.
For instance, one could use the dates of employment
to discriminate between multiple edges modeling
recurring employments of a LinkedIn user at the same company.

\lstset{basicstyle=\fontsize{8.5}{7.5}\ttfamily,style=GSQL}  
\begin{lstlisting}
CREATE DIRECTED EDGE Employment
   (FROM Company, TO Person, start DATE, (*@{\tt end}@*) DATE)
   (*@{\tt\textbf{DISCRIMINATOR}}@*) (start)
\end{lstlisting}

\myparagraph{Reverse Edges}
The GSQL data model includes the concept of edges being inverses
of each other, analogous to the notion of
inverse relationships from the ODMG ODL standard~\cite{odmg}.

Consider a graph of fund transfers 
between bank accounts, with a directed Debit edge from account
A to account B signifying the debiting of A in favor of B
(the amount and timestamp would be modeled as edge attributes).
The debiting action corresponds to a crediting action in the opposite
sense, from B to A.
If the application needs to explicitly keep track of both credit and
debit vocabulary terms, a natural modeling consists in introducing
for each Debit edge a reverse Credit edge for the same endpoints,
with both edges sharing the values of the attributes, as in the following
example:

\lstset{basicstyle=\fontsize{8.5}{7.5}\ttfamily,style=GSQL}  
\begin{lstlisting}
CREATE VERTEX Account
   (number int PRIMARY KEY, balance FLOAT, ...)

CREATE DIRECTED EDGE Debit
   (FROM Account, TO Account, amount float, ...)
   WITH REVERSE EDGE Credit
\end{lstlisting}

\section{GSQL DML}
\label{sec:dml}

The guiding principle behind the design of GSQL was to
facilitate adoption by SQL programmers
while simultaneously flattening the learning curve for novices,
especially for adopters of the BSP programming paradigm~\cite{bsp}.

To this end, GSQL's design starts from SQL, extending its syntax and
semantics parsimoniously, i.e.
avoiding the introduction of new keywords for concepts that already have an SQL counterpart.
We first summarize the key additional primitives before detailing them.

\myparagraph{Graph Patterns in the \from\ Clause}
GSQL extends SQL's \from\ clause to allow the specification of {\em patterns}.
Patterns specify constraints on paths in the graph, and they 
also contain variables, bound to vertices or edges of the matching paths.
In the remaining query clauses, these variables are treated just like standard SQL tuple
variables.

\myparagraph{Accumulators}
The data found along a path matched by a pattern
can be collected and aggregated into {\em accumulators}.
Accumulators support multiple simultaneous aggregations of the same data according to
distinct grouping criteria.
The aggregation results can be distributed across vertices, to support multi-pass
and, in conjunction with loops, even iterative graph algorithms implemented in MPP fashion.

\myparagraph{Loops}
GSQL includes control flow primitives, in particular loops, which are essential 
to support standard iterative graph analytics
(e.g. PageRank~\cite{pagerank}, shortest-paths~\cite{gibbons-graph-book},
weakly connected components~\cite{gibbons-graph-book}, recommender systems, etc.).

\myparagraph{Direction-Aware Regular Path Expressions (\darpe s)}\\
GSQL's \from\ clause patterns contain path expressions that
specify constrained reachability queries in the graph.
GSQL path expressions start from the de facto standard of
two-way regular path expressions~\cite{crpqi} which is the culmination
of a long line of works on graph query languages, including reference languages
like WebSQL~\cite{websql}, StruQL~\cite{struql} and Lorel~\cite{lorel}.
Since two-way regular path expressions were developed for directed graphs,
GSQL extends them to support both directed and undirected
edges in the same graph.
We call the resulting path expressions
{\em Direction-Aware Regular Path Expressions (\darpe s)}.

\subsection{Graph Patterns in the \from\ Clause}
GSQL's \from\ clause extends SQL's basic \from\ clause syntax to also allow
atoms of general form\\

$<$GraphName$>$ AS? $<$pattern$>$\\

\noindent
where the AS keyword is optional, $<$GraphName$>$ is the name of a graph, and
<pattern> is a pattern given by a regular path expression with variables.

This is in analogy to standard SQL, in which a FROM clause atom\\

$<$TableName$>$ AS? $<$Alias$>$\\

\noindent
specifies a collection (a bag of tuples) to the left of the AS keyword
and introduces an alias to the right. This alias can
be viewed as a simple pattern that introduces a single tuple variable.
In the graph setting, the collection is the graph and the pattern may introduce
several variables.
We show more complex patterns in Section~\ref{sec:darpes}
but illustrate first with the following simple-pattern example.

\begin{myexample}{Seamless Querying of Graphs and Relational Tables}
  \label{ex:seamless}
Assume Company ACME maintains a human resource database stored in an RDBMS
containing a relational table “Employee”. It also has access to the
“LinkedIn" graph from Example~\ref{ex:ddl} containing the professional network of LinkedIn users.

The query in Figure~\ref{fig:seamless}
joins relational HR employee data with LinkedIn graph data to find the
employees who have made the most LinkedIn connections outside the company since 2016:

\begin{figure*}
  \center
  \lstset{basicstyle=\fontsize{8.5}{7.5}\ttfamily,style=GSQL}
  
  \begin{lstlisting}
    SELECT e.email, e.name, count (outsider)
    FROM   Employee AS e,
           LinkedIn AS Person: p -(Connected: c)- Person: outsider
    WHERE  e.email = p.email and
           outsider.currentCompany NOT LIKE 'ACME' and
           c.since >= 2016
    GROUP BY e.email, e.name
  \end{lstlisting}
  
  \caption{Query for Example~\ref{ex:seamless}, Joining Relational Table and Graph}
  \label{fig:seamless}
\end{figure*}

Notice the pattern Person:p -(Connected:c)- Person:outsider
to be matched against the “LinkedIn“ graph.
The pattern variables are “p”, “c” and “outsider”, binding respectively to a
“Person” vertex, a “Connected” edge and a “Person” vertex.
Once the pattern is matched, its variables can be used just like standard SQL
tuple aliases. Notice that neither the WHERE clause nor the SELECT clause
syntax discriminate among aliases, regardless of whether they range over tuples,
vertices or edges.

The lack of an arrowhead accompanying the edge subpattern
-(Connected: c)-
requires the matched “Connected” edge to be undirected.
\end{myexample}

To support cross-graph joins, the FROM clause allows the mention of multiple
graphs, analogously to how the standard SQL FROM clause may mention multiple tables.

\begin{myexample}{Cross-Graph and -Table Joins}
  \label{ex:cross-join}
  Assume we wish to gather information on employees,
  including how many tweets about their company and how many LinkedIn connections they have.
  The employee info resides in a relational table “Employee”, the LinkedIn data is in
  the graph named “LinkedIn” and the tweets are in the graph named “Twitter”.
  The query is shown in Figure~\ref{fig:cross-join}.
  Notice the join across the two graphs and the relational table.

  \begin{figure*}
    \center
    \lstset{basicstyle=\fontsize{8.5}{7.5}\ttfamily,style=GSQL}

    \begin{lstlisting}
      SELECT e.email, e.name, e.salary, count (other), count (t)
      FROM   Employee AS e,
             LinkedIn AS Person: p -(Connected)- Person: other,
             Twitter AS User: u -(Posts>)- Tweet: t
      WHERE  e.email = p.email and p.email = u.email and
             t.text CONTAINS e.company
      GROUP BY e.email, e.name, e.salary
    \end{lstlisting}

    \caption{Query for Example~\ref{ex:cross-join}, Joining Across Two Graphs}
    \label{fig:cross-join}
  \end{figure*}
  Also notice the arrowhead in the edge subpattern
  ranging over the Twitter graph, $-(Posts>)-$, which
  matches only directed edges of type “Posts“,
  pointing from the “User” vertex to the “Tweet” vertex.
\end{myexample}

\subsection{Accumulators}
We next introduce the concept of accumulators ,
i.e. data containers that store an internal value and take
inputs that are aggregated into this internal value using a binary operation.
Accumulators support the concise specification of multiple simultaneous
aggregations by distinct grouping criteria, and the computation of vertex-stored
side effects to support multipass and iterative algorithms.
\AD{
As detailed in Section~\ref{sec:mpp-interpretation}, they are also crucial in
enabling developers to preserve their BSP and Map/Reduce programming style
when writing GSQL queries.
}

The accumulator abstraction was introduced in the GreenMarl
system~\cite{green-marl} and it was adapted as
high-level first-class citizen in GSQL to distinguish among two flavors:
\begin{itemize}
\item
  {\em Vertex accumulators} are attached to vertices,
  with each vertex storing its own local accumulator instance.
  They are useful in aggregating data encountered during the traversal of path patterns
  and in storing the
  result distributed over the visited vertices.
\item
 {\em Global accumulators} have a single instance and
 are useful in computing global aggregates.
\end{itemize}

Accumulators are polymorphic, being parameterized by the type of the internal value $V$,
the type of the inputs $I$, and the binary {\em combiner} operation
$$\oplus: V \times I \rightarrow V.$$

Accumulators implement two assignment
operators. Denoting with $a.\val$ the internal value of accumulator $a$,
\begin{itemize}
\item
  $a\ = i$ sets $a.\val$ to the provided input $i$;
\item
  $a\ \incr i$ aggregates the input $i$ into $acc.\val$ using the combiner,
  i.e. sets $a.\val$ to $a.\val \oplus i$.
\end{itemize}

For a
comprehensive documentation on GSQL accumulators, see the developer’s guide at
\myhref{http://docs.tigergraph.com}{http://docs.tigergraph.com}.
Here, we explain accumulators by example.

\begin{myexample}{Multiple Aggregations by Distinct Grouping Criteria}
  \label{ex:multi-agg}
    Consider a graph named “SalesGraph” in which the sale of a product
    $p$ to a customer $c$ is modeled by a directed  “Bought”-edge
    from the “Customer”-vertex modeling $c$ to the “Product”-vertex modeling $p$.
    The number of product units bought, as well as the discount at
    which they were offered are recorded as attributes of the edge.
    The list price of the product is stored as attribute of the corresponding “Product”
    vertex.
    
    We wish to simultaneously compute the sales revenue per product from the “toy” category,
    the toy sales revenue per customer, and the overall total toy sales revenue.~\footnote{
      Note that writing this query in standard SQL is cumbersome.
      It requires performing two GROUP BY operations, one
      by customer and one by product. Alternatively, one can use the window functions'
      OVER - PARTITION BY clause, that can perform
      the groupings independently but whose output repeats the customer revenue for each
      product bought by the customer, and the product revenue for each customer buying the product.
      Besides yielding an unnecessarily large result, this solution then
      requires two post-processing SQL queries to separate the two aggregates.
    }

    We define a vertex accumulator type for each kind of revenue.       
    The revenue for toy product $p$ will be
    aggregated at the vertex modeling $p$ by vertex accumulator {\tt revenuePerToy},
    while the revenue for customer $c$ will be aggregated at the vertex modeling $c$ by
    the vertex accumulator {\tt revenuePerCust}.
    The total toy sales revenue will be aggregated in a global accumulator called
    {\tt totalRevenue}.
    With these accumulators, the multi-grouping
    query is concisely expressible (Figure~\ref{fig:multi-agg}).

    \begin{figure*}
      \center
      \lstset{basicstyle=\fontsize{8.5}{7.5}\ttfamily,style=GSQL}
      \begin{lstlisting}
        WITH
          SumAccum<float> @revenuePerToy, @revenuePerCust, @@totalRevenue
        BEGIN
          
          SELECT c
          FROM   SalesGraph AS Customer: c -(Bought>: b)- Product:p
          WHERE  p.category = 'toys'
          ACCUM  float salesPrice = b.quantity * p.listPrice * (100 - b.percentDiscount)/100.0,
                 c.@revenuePerCust += salesPrice,
                 p.@revenuePerToy += salesPrice,
                 @@totalRevenue += salesPrice;

        END
      \end{lstlisting}
      \caption{Multi-Aggregating Query for Example~\ref{ex:multi-agg}}
      \label{fig:multi-agg}
    \end{figure*}
    Note the definition of the accumulators using the WITH clause in the spirit of
    standard SQL definitions. Here,\\ {\tt SumAccum<float>}
    denotes the type of accumulators that hold an internal floating point scalar value
    and aggregate inputs using the floating point addition operation.
    Accumulator names prefixed by a single @ symbol denote vertex accumulators
    (one instance per vertex)  while accumulator names prefixed by
    @@ denote a global accumulator (a single shared instance).

    Also note the novel \accum\ clause, which specifies the generation of inputs to
    the accumulators. Its first line introduces a local variable “salesPrice”, whose value
    depends on attributes found in both the “Bought” edge and the “Product” vertex.
    This value is aggregated into each accumulator using the “+=” operator.
    {\tt c.@revenuePerCust} refers to the vertex accumulator instance located at the
    vertex denoted by vertex variable $c$.
\end{myexample}

\myparagraph{Multi-Output SELECT Clause}
GSQL's accumulators allow the simultaneous specification of multiple
aggregations of the same data. To take full advantage of this capability,
GSQL complements it with the ability to concisely
specify simultaneous outputs into multiple tables for data obtained by the
same query body. This can be thought of as evaluating multiple independent
SELECT clauses.

\begin{myexample}{Multi-Output SELECT}
\label{ex:multi-output}
  While the query in Example~\ref{ex:multi-agg}
  outputs the customer vertex ids, in that example we were interested in
  its side effect of annotating vertices with the aggregated revenue values and of
  computing the total revenue. If instead we wished to create two tables,
  one associating customer names with their revenue, and one associating toy names
  with theirs, we would employ GSQL's {\em multi-output SELECT clause} as follows
  (preserving the \from, \where\ and \accum\ clauses of Example~\ref{ex:multi-agg}).
  
  \lstset{basicstyle=\fontsize{8.5}{7.5}\ttfamily,style=GSQL}
  \begin{lstlisting}
  SELECT c.name, c.@revenuePerCust INTO PerCust;
         t.name, t.@revenuePerToy INTO PerToy
  \end{lstlisting}
  Notice the semicolon, which separates the
  two simultaneous outputs.
\end{myexample}

\myparagraph{Semantics}
The semantics of GSQL queries can be given in a
declarative fashion analogous to SQL semantics:
for each distinct match of the FROM clause pattern that satisfies the WHERE
clause condition, the ACCUM clause is executed precisely once.
After the ACCUM clause executions complete, the multi-output SELECT clause
executes each of its semicolon-separated individual fragments independently,
as standard SQL clauses.
Note that we do
not specify the order in which matches are found and consequently the order of ACCUM clause
applications. We leave this to the engine implementation to support optimization.
The result is well-defined (input-order-invariant) whenever the accumulator’s binary
aggregation operation $\oplus$ is commutative and associative. This is
certainly the case for addition, which is used in Example~\ref{ex:multi-agg},
and for most of GSQL's built-in accumulators. 

\myparagraph{Extensible Accumulator Library}
GSQL offers a list of built-in accumulator types. TigerGraph’s experience
with the deployment of GSQL has yielded the short list from
Section~\ref{sec:built-in-accums}, that covers most use cases we
have encountered in customer engagements.
In addition, GSQL allows users to define their own accumulators by implementing
a simple C++ interface that declares the binary combiner operation $\oplus$
used for aggregation of inputs into the stored value.
This leads to an extensible query language, facilitating the development of accumulator
libraries.

\myparagraph{Accumulator Support for Multi-pass Algorithms}
The scope of the accumulator declaration may cover a sequence of query blocks,
in which case the accumulated values computed by a block can be read
(and further modified) by subsequent blocks, thus achieving powerful composition effects.
These are particularly useful in multi-pass algorithms.

\begin{myexample}{Two-Pass Recommender Query}
  \label{ex:recommender}
  Assume we wish to write a simple toy recommendation system for a customer $c$ given
  as parameter to the query. The recommendations are ranked in the classical manner:
  each recommended toy’s rank is a weighted sum of the likes by other customers.
  
  Each like by an other customer $o$ is weighted by the similarity of $o$ to customer
  $c$. In this example, similarity is the standard log-cosine similarity~\cite{log-cosine},
  which reflects how many toys customers $c$ and $o$ like in common.
  Given two customers x and y, their log-cosine similarity is defined as
  $log(1 + \mathit{count\ of\ common\ likes\ for\ } x \mathit{\ and\ } y)$.

  \begin{figure*}
    \center
    \lstset{basicstyle=\fontsize{8.5}{7.5}\ttfamily,style=GSQL}
    \begin{lstlisting}
    CREATE QUERY TopKToys (vertex<Customer> c, int k) FOR GRAPH SalesGraph {
      
      SumAccum<float> @lc, @inCommon, @rank;
      
      SELECT DISTINCT o INTO OthersWithCommonLikes
      FROM   Customer:c -(Likes>)- Product:t -(<Likes)- Customer:o
      WHERE  o <> c and t.category = 'Toys'
      ACCUM  o.@inCommon += 1
      POST_ACCUM o.@lc = log (1 + o.@inCommon);

      SELECT t.name, t.@rank AS rank INTO Recommended
      FROM   OthersWithCommonLikes:o -(Likes>)- Product:t
      WHERE  t.category = 'Toy' and c <> o
      ACCUM  t.@rank += o.@lc
      ORDER BY t.@rank DESC
      LIMIT  k;

      RETURN Recommended;
    }    
    \end{lstlisting}
    \caption{Recommender Query for Example~\ref{ex:recommender}}
    \label{fig:recommender}
  \end{figure*}

  The query is shown in Figure~\ref{fig:recommender}.
  The query header declares the name of the query
  and its parameters (the vertex of type ``Customer'' $c$, and the integer value $k$
  of desired recommendations). The header also declares the graph
  for which the query is meant, thus freeing the programmer from
  repeating the name in the \from\ clauses. Notice also that the accumulators
  are not declared in a WITH clause. In such cases, the GSQL convention is that the
  accumulator scope spans all query blocks.
  Query {\tt TopKToys} consists of two blocks.
  
  The first query block computes for each other customer $o$ their log-cosine
  similarity to customer $c$, storing it in $o$’s vertex accumulator {\tt @lc}.
  To this end, the \accum\ clause first counts the toys liked in common by
  aggregating for each such toy the value $1$ into $o$’s vertex accumulator
  {\tt @inCommon}. The \paccum\ clause then computes the logarithm and stores
  it in $o$’s vertex accumulator {\em @lc}.

  Next, the second query block computes the rank of each toy $t$ by adding up
  the similarities of all other customers $o$ who like $t$.
  It outputs the top $k$ recommendations into table {\tt Recommended}, which is
  returned by the query.

  Notice the input-output
  composition due to the second query block’s \from\ clause referring to the set of vertices
  {\tt OthersWithCommonLikes} (represented as a single-\\column table)
  computed by the first query block. Also notice the side-effect
  composition due to the second block’s \accum\ clause referring to the
  {\tt @lc} vertex accumulators computed by the first block. Finally,
  notice how the SELECT clause outputs vertex accumulator values ({\tt t.@rank})
  analogously to how it outputs vertex attributes ({\tt t.name}).
\end{myexample}
Example~\ref{ex:recommender} introduces the \paccum\ clause, which is a convenient
way to post-process accumulator values after the \accum\ clause
finishes computing their new aggregate value.

\subsection{Loops}
GSQL includes a while loop primitive, which, when combined with
accumulators, supports iterative graph algorithms. We illustrate
for the classic PageRank~\cite{pagerank} algorithm.

\begin{myexample}{PageRank}
\label{ex:pagerank}
Figure~\ref{fig:pagerank} shows a GSQL query implementing a simple
version of PageRank.

\begin{figure*}
\center
\lstset{basicstyle=\fontsize{8.5}{7.5}\ttfamily,style=GSQL}
\begin{lstlisting}
CREATE QUERY PageRank (float maxChange, int maxIteration, float dampingFactor) {

  MaxAccum<float> @@maxDifference;        //(*@ max @*)score change(*@ in @*)an iteration
  SumAccum<float> @received_score;        //(*@ sum @*)of scores received(*@ from @*)neighbors
  SumAccum<float> @score = 1;             // initial score for every(*@ vertex @*)is 1.

  AllV = {Page.*};                        // start with(*@ all @*)vertices of type Page
  
  WHILE @@maxDifference > maxChange LIMIT maxIteration DO
     @@maxDifference = 0;

     S = SELECT      v
         FROM        AllV:v -(LinkTo>)- Page:n
         ACCUM       n.@received_score += v.@score/v.outdegree()
         POST-ACCUM  v.@score = 1-dampingFactor + dampingFactor * v.@received_score,
                     v.@received_score = 0,
                     @@maxDifference += abs(v.@score - v.@score');
  END;
}
\end{lstlisting}
\caption{PageRank Query for Example~\ref{ex:pagerank}}
\label{fig:pagerank}
\end{figure*}

Notice the while loop that runs a maximum number of iterations provided
as parameter {\tt maxIteration}.
Each vertex $v$ is equipped with a {\tt @score}
accumulator that computes the rank at each iteration,
based on the current score at $v$ and the sum of fractions of previous-iteration scores  
of $v$'s neighbors (denoted by vertex variable $n$). {\tt v.@score'} refers to the value of
this accumulator at the previous iteration.

According to the \accum\ clause, at every iteration 
each vertex $v$ contributes to its neighbor $n$'s score a fraction of $v$'s
current score, spread over $v$'s outdegree.
The score fractions contributed by the neighbors are
summed up in the vertex accumulator {\tt @received\_score}.

As per the \paccum\ clause, once the sum of score fractions
is computed at $v$, it is combined linearly with $v$'s current score based on the
parameter {\tt dampingFactor}, yielding a new score for $v$.

The loop terminates early if the maximum difference over all vertices
between the previous iteration's score (accessible as {\tt v.@score'}) and the new
score (now available in {\tt v.@score}) is within a threshold given by parameter {\tt maxChange}.
This maximum is computed in the {\tt @@maxDif\-ference} global accumulator, which receives as
inputs the absolute differences computed by the \paccum\ clause instantiations for every value
of vertex variable $v$.
\end{myexample}

\subsection{\darpe s}\label{sec:darpes}
We follow the tradition instituted
by a line of classic work on querying graph (a.k.a. semi-structured)
data which yielded such reference query languages as WebSQL~\cite{websql},
StruQL~\cite{struql} and Lorel~\cite{lorel}.
Common to all these languages is a primitive that allows the programmer
to specify traversals along paths whose structure is constrained by a regular
path expression.

{\em Regular path expressions (RPEs)} are regular expressions over the alphabet of
edge types. They conform to the context-free grammar
\begin{eqnarray*}
  \mathit{rpe} & \rightarrow &
  \_ \ |\ \mathit{EdgeType}\ |\ '('\ \mathit{rpe}\ ')' | \mathit{rpe}\ '*'\ bounds?\ \\
  & | & rpe\ '.'\ rpe\ |\ rpe\ '|'\ rpe\\
\mathit{bounds} & \rightarrow & N?\ '..'\ N?
\end{eqnarray*}
where {\em EdgeType} and $N$ are terminal symbols representing
respectively the name of an edge type and a natural number.
The wildcard symbol “\_” denotes any edge type, “.” denotes the concatenation of
its pattern arguments, and “$|$” their disjunction.
The ‘*’ terminal symbol is the standard Kleene star specifying several (possibly $0$
or unboundedly many) repetitions of its RPE argument.
The optional bounds can specify a minimum and a maximum
number of repetitions (to the left and right of the “..” symbol, respectively).

A path $p$ in the graph is said to satisfy an RPE $R$ if the sequence of edge types
read from the source vertex of $p$ to the target vertex of $p$ spells out a word in the language
accepted by $R$ when interpreted as a standard regular expression over the alphabet of
edge type names.

\myparagraph{\darpe s}
Since GSQL's data model allows for the existence of both directed and undirected edges
in the same graph,
we refine the RPE formalism, proposing {\em Direction-Aware RPEs (DARPEs)}.
These allow one to also specify the orientation of directed edges in the path.
To this end, we extend the alphabet to include for
each edge type $E$ the symbols
\begin{itemize}
\item
  $E$, denoting a hop along an undirected $E$-edge,
\item
  $E\!\!>$, denoting a hop along an outgoing $E$-edge (from source to target vertex), and
\item
  $<\!\!E$, denoting a hop along an incoming $E$-edge (from target to source vertex).
\end{itemize}
Now the notion of satisfaction of a DARPE by a path extends classical RPE satisfaction
in the natural way.

DARPEs enable the free mix of edge directions in regular path expressions.
For instance, the pattern
$$
E\!\!>.(F\!\!>|<\!\!G)*.<\!\!H.J
$$
matches paths starting with a hop along an outgoing $E$-edge,
followed by a sequence of zero or more hops along either outgoing $F$-edges
or incoming $G$-edges, next by a hop along an incoming $H$-edge
and finally ending in a hop along an undirected $J$-edge.

\subsubsection{\darpe\ Semantics}\ \\
A well-known semantic issue arises from the tension between RPE expressivity
and well-definedness. Regarding expressivity, applications need to sometimes specify
reachability in the graph via RPEs comprising unbounded (Kleene) repetitions of a path shape
(e.g. to find which target users are influenced by source users on Twitter,
we seek the paths  connecting users directly or indirectly
via a sequence of tweets or retweets).
Applications also need to compute various aggregate statistics over the graph,
many of which are multiplicity-sensitive (e.g. count, sum, average). Therefore,
pattern matches must preserve multiplicities, being interpreted under bag semantics.
That is, a pattern :s -(RPE)- :t
should have as many matches of variables $(s,t)$ to a given pair
of vertices $(n_1,n_2)$ as there are distinct paths from
$n_1$ to $n_2$ satisfying the RPE. In other words, the count of these paths
is the multiplicity of the $(n_1,n_2)$ in the bag of matches.

The two requirements conflict with well-definedness:
when the RPE contains Kleene stars, cycles in the
graph can yield an infinity of distinct paths satisfying the RPE
(one for each number of times the path wraps around the cycle),
thus yielding infinite multiplicities in the query output.
Consider for example the pattern\\
{\em Person: p1 -(Knows$>$*)- Person: p2}
in a social network with cycles involving the “Knows” edges.

\myparagraph{Legal Paths}
Traditional solutions limit the kind of paths that are considered legal,
so as to yield a finite number in any graph.
Two popular approaches allow only paths with non-repeated vertices/edges.~\footnote{
  Gremlin's~\cite{gremlin} default semantics allows all unrestricted paths (and therefore possibly
  non-terminating graph traversals), but virtually all the documentation
  and tutorial examples involving unbounded traversal use non-repeated-vertex semantics
  (by explicitly invoking a built-in {\tt simplePath} predicate). By default,
  Cypher~\cite{neo4j} adopts the non-repeated-edge semantics.
}
However under these definitions of path legality the evaluation of RPEs is in general
notoriously intractable:
even checking existence of legal paths that satisfy the RPE (without counting them)
has worst-case NP-hard data complexity
(i.e. in the size of the graph \cite{MW95,LMV16}).
As for the process of counting such paths, it is \#P-complete. 
This worst-case complexity does not scale to large graphs.

In contrast, GSQL adopts the {\em all-shortest-paths} legality criterion.
That is, among the paths from $s$ to $t$ satisfying a given
\darpe, GSQL considers legal all the shortest ones.
Checking existence of a shortest path that satisfies a \darpe, and even counting
all such shortest paths is tractable (has polynomial data complexity).

For completeness, we recall here also the semantics adopted by the SparQL
standard~\cite{sparql-standard}
(SparQL is the W3C-standardized query language for RDF graphs):
SparQL regular path expressions that are Kleene-starred are interpreted
as boolean {\em tests} whether such a path exists, without counting
the paths connecting a pair of endpoints.
This yields a multiplicity of $1$ on the pair of path endpoints,
which does not align with our goal of maintaining bag semantics
for aggregation purposes.

\eat{
This design choice is a twist on the
semantics advocated in~\cite{AAB+17} to guarantee tractability,
namely the non-deterministic choice of one among the shortest paths.
This would however defeat the purpose of multiplicity-aware aggregation
and therefore GSQL (like other languages in use today, for instance Cypher and Gremlin)
does not conform to the choose-one semantics.
}

\begin{myexample}{Contrasting Path Semantics}
\label{ex:paths-semantics}

  \begin{figure}
    \centering
    \includegraphics[width=0.4\textwidth,trim={3cm 4cm 4cm 4cm},clip=true]{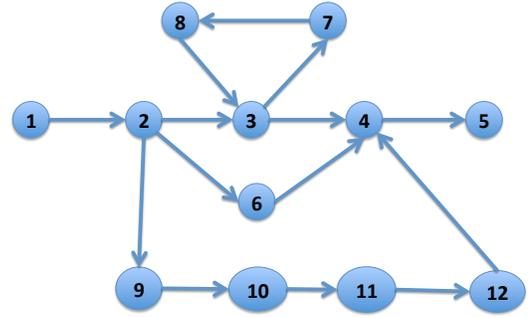}
\caption{Graph $G_1$ for Example~\ref{ex:paths-semantics}}
\label{fig:g1}
\end{figure} 

  To contrast the various path legality flavors, consider the graph $G_1$
  in Figure~\ref{fig:g1}, assuming that
  all edges are typed “E”.
  Among all paths from source vertex 1 to target vertex 5 that satisfy the \darpe\
  “$E\!\!>*$”,
  there are
\begin{itemize}
  \item
    Infinitely many unrestricted paths, depending on how many times they wrap around
    the 3-7-8-3 cycle;
  \item
    Three non-repeated-vertex paths (1-2-3-4-5, 1-2-6-4-5, and 1-2-9-10-11-12-4-5);
  \item
    Four non-repeated-edge paths (1-2-3-4-5, 1-2-6-4-5, 1-2-9-10-11-12-4-5, and
    1-2-3-7-8-3-4-5);
  \item
    Two shortest paths (1-2-3-4-5 and 1-2-6-4-5).
\end{itemize}

Therefore, pattern $:s -(E\!\!>*)- :t$
will return the binding $(s \mapsto 1,t \mapsto 5)$
with multiplicity 3, 4, or 2 under the non-repeated-vertex,
non-repeated-edge respectively shortest-path legality criterion.
In addition, under SparQL semantics, the multiplicity is 1.

\begin{figure}
  \centering
  \includegraphics[width=0.35\textwidth,trim={6cm 9cm 6cm 6cm},clip=true]{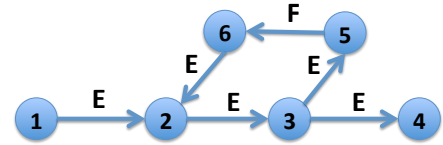}
  \caption{Graph $G_2$ for Example~\ref{ex:paths-semantics}}
  \label{fig:g2}
\end{figure} 

While in this example the shortest paths are a subset of the non-repeated-vertex paths,
which in turn are a subset of the non-repeated-edge paths, this inclusion does not hold
in general, and the different classes are incomparable.
Consider Graph $G_2$ from Figure~\ref{fig:g2}, and the pattern
$$
:s -(E\!\!>*. F\!\!>. E\!\!>*)- :t
$$
and note that it does not match any path from vertex 1 to vertex 4
under non-repeated vertex or edge semantics, while it does match one such path
under shortest-path semantics (1-2-3-5-6-2-3-4).
\end{myexample}

\subsection{Updates}
GSQL supports vertex-, edge-  as well as attribute-level modifications
(insertions, deletions and updates), with a syntax inspired by SQL
(detailed in the online documentation at
\myhref{https://docs.tigergraph.com/dev/gsql-ref}{https://docs.tigergraph.com/dev/gsql-ref}).
 
\

\section{GSQL Evaluation Complexity}
\label{sec:complexity}

Due to its versatile control flow and user-defined accumulators,
GSQL is evidently a Turing-complete language and therefore it does not make
sense to discuss query evaluation complexity for the unrestricted language.

However, there are natural restrictions for which such considerations are meaningful,
also allowing comparisons to other languages: if we rule out loops, accumulators and
negation, but allow \darpe s, we are left essentially with a language that corresponds
to the class of aggregating conjunctive two-way regular path queries~\cite{aggregating-crpqs}.

The evaluation of queries in this restricted GSQL language fragment
has polynomial data complexity.
This is due to the shortest-path semantics
and to the fact that, in this GSQL fragment, one need not {\em enumerate}
the shortest paths as it suffices to {\em count} them in order to maintain the
multiplicities of bindings. While enumerating paths (even only shortest)
may lead to exponential result size in the size of the input graph,
counting shortest paths is tractable (i.e. has polynomial data complexity).

\myparagraph{Alternate Designs Causing Intractability}
If we change the path legality criterion (to the defaults from languages such as Gremlin and Cypher),
tractability is no longer given. An additional obstacle to tractability is the primitive
of binding a variable to the entire path (supported in both Gremlin and Cypher
but not in GSQL), which may lead to results of size exponential in the input graph size,
even under shortest-paths semantics.

\myparagraph{Emulating the Intractable (and Any Other) Path Legality Variants}
GSQL accumulators and loops constitute a powerful combination that
can implement both flavors of the intractable path semantics (non-repeated vertex or edge
legality), and beyond, {\em as syntactic sugar}. That is, without requiring further
extensions to GSQL.

To this end, the programmer can split the pattern into a sequence of query blocks,
one for each single hop in the path, explicitly materializing legal paths in
vertex accumulators whose contents is transferred
from source to target of the hop. In the process, the paths materialized in the source
vertex accumulator are extended with the current hop, and the extensions are stored
in the target's vertex accumulator.

For instance, a query block with the \from\ clause  pattern
S:s -(E.F)- T:t is implemented by a sequence of two blocks,
one with \from\ pattern S:s-(E)- \_:x, followed by one with \from\ pattern
\_:x -(F)- T:t. Similar transformations can eventually translate arbitrarily complex
\darpe s into sequences of query blocks whose patterns specify single hops
(while loops are needed to implement Kleene repetitions).

For each single-hop block, whenever a pattern S:s -(E)- T:t
is matched, the paths leading to the vertex denoted by $s$ are kept in a vertex
accumulator, say {\tt s.@paths}. These paths are extended with an additional hop to
the vertex denoted by $t$, the extensions are filtered by the desired legality
criterion, and the qualifying extensions are added to {\tt t.@paths}.

Note that this translation can be performed automatically, thus allowing
programmers to specify
the desired evaluation semantics
{\em on a per-query basis} and even
{\em on a per-\darpe\ basis} within the same query.
The current GSQL release does not include this automatic translation,
but future releases might do so given sufficient customer demand (which is yet to
materialize). Absent such demand, we believe that there is value in making
programmers work harder (and therefore think harder whether they really want)
to specify intractable semantics.

\section{BSP Interpretation of GSQL}\label{sec:mpp-interpretation}
\label{sec:mpp-intepretation}

GSQL is a high-level language that admits a semantics that is
compatible with SQL and hence declarative and agnostic of the
computation model. This presentation appeals to SQL experts as well as to novice
users who wish to abstract away computation model details.

However, GSQL was designed from the get-go for full compatibility
with the classical programming paradigms for BSP programming embraced by
developers of graph analytics and more generally, NoSQL Map/Reduce
jobs. We give here alternate interpretations to
GSQL, which allow developers to preserve their
Map/Reduce or graph BSP mentality while gaining the benefits of high-level
specification when writing GSQL queries.

\subsection{Think-Like-A-Vertex/Edge Interpretation}
The TigerGraph graph can be viewed both as a data model and a computation model.
As a computation model, the programming abstraction offered by TigerGraph
integrates and extends the two classical graph programming paradigms of
think-like-a-vertex
(a la Pregel~\cite{pregel}, GraphLab~\cite{graphlab} and Giraph~\cite{giraph})
and think-like-an-edge (PowerGraph~\cite{powergraph}).
A key difference between TigerGraph's computation model and that
of the above-mentioned systems is that these allow the programmer to specify only one
function that uniformly aggregates messages/inputs received at a vertex, while in GSQL
one can declare an unbounded number of accumulators of different types.

Conceptually, each vertex or edge acts as a parallel unit of storage and computation
simultaneously, with the graph constituting a massively parallel computational mesh.
Each vertex and edge can be associated with a {\em compute function},
specified by the \accum\ clause (vertex compute functions can also be
specified by the \paccum\ clause).

The compute function is instantiated for each edge/vertex,
and the instantiations execute in parallel, generating accumulator inputs ({\em acc-inputs}
for short).
We refer to this computation as the {\em acc-input generation phase}.
During this phase, the compute function instantiations work under {\em snapshot semantics}:
the effect of incorporating generated inputs into accumulator values is not visible
to the individual executions until all of them finish. This guarantees that all
function instantiations can execute in parallel, starting from the same accumulator
snapshot, without interference as there are no dependencies between them.

Once the acc-input generation phase completes,
the {\em aggregation phase} ensues. Here, inputs are aggregated into
their accumulators using the appropriate binary combiners. Note that vertex accumulators
located at different vertices can work in parallel, without interfering with each other.

This semantics requires synchronous execution, with the
aggregation phase waiting at a {\em synchronization barrier} until the acc-input generation
phase completes (which in turn waits at a synchronization barrier until the
preceding query block's aggregation phase completes, if such a block exists).
It is therefore an instantiation of the Bulk-Synchronous-Parallel processing
model~\cite{bsp}.

\subsection{Map/Reduce Interpretation}
GSQL also admits an interpretation that appeals to NoSQL practitioners who
are not specialized on graphs and are trained on the classical Map/Reduce paradigm.

This interpretation is more evident on a normal form of GSQL queries,
in which the \from\ clause contains a single pattern that specifies a one-hop
\darpe: S:s -(E:e)- T:t, with E a single edge type or a disjunction thereof.
It is easy to see that all GSQL queries can be normalized in this way (though such
normalization may involve the introduction of accumulators to implement the flow
of data along multi-hop traversals, and of loops to implement Kleene repetitions).
Once normalized, all GSQL queries can be interpreted as follows.

In this interpretation, the \from, \where\ and \accum\ clauses together
specify an {\em edge map (EM)} function, which is mapped over all edges in the graph.
For those edges that satisfy the \from\ clause pattern and the \where\ condition,
the EM function evaluates the \accum\ clause and
outputs a set of key-value pairs, where the key identifies an accumulator
and the value an acc-input. The sending to accumulator $A$ of all inputs destined for $A$
corresponds to the {\em shuffle phase} in the classical map/reduce paradigm,
with $A$ playing the role of a {\em reducer}. Since for vertex accumulators the
aggregation happens at the vertices, we refer to this phase as the
{\em vertex reduce (VR)} function.

Each GSQL query block thus specifies a variation of Map/Reduce jobs
which we call {\em EM/VR jobs}. As opposed to standard map/reduce jobs,
which define a single map/reduce function pair,
a GSQL EM/VR job can specify several reducer types at once
by defining multiple accumulators.

\section{Conclusions}
\label{sec:conclusions}

TigerGraph's design is pervaded by MPP-awareness
in all aspects, ranging from
(i)
the native storage with its customized partitioning, compression and layout schemes
to
(ii)
the execution engine architected as a multi-server cluster that minimizes cross-boundary communication
and exploits multi-threading within each server,
to
(iii)
the low-level graph BSP programming abstractions and
to
(iv)
the high-level query language admitting BSP interpretations.

This design yields noteworthy scale-up and scale-out performance whose experimental
evaluation is reported in Appendix~\ref{sec:benchmark} for a benchmark that we made avaiable
on GitHub for full reproducibility~\footnote{\myhref{https://github.com/tigergraph/ecosys/tree/benchmark/benchmark/tigergraph}{https://github.com/tigergraph/ecosys/tree/benchmark/benchmark/tigergraph}}. We used this benchmark also for comparison with other leading graph database
systems\\ (ArangoDB~\cite{arangodb}, Azure CosmosDB~\cite{cosmosdb}, JanusGraph~\cite{janusgraph},
Neo4j~\cite{neo4j}, Amazon Neptune~\cite{neptune}), against which
TigerGraph compares favorably, as detailed in an online
report~\footnote{\myhref{https://www.tigergraph.com/benchmark/}{https://www.tigergraph.com/benchmark/}}.
The scripts for these systems are also published on
GitHub~\footnote{\myhref{https://github.com/tigergraph/ecosys/tree/benchmark/benchmark}{https://github.com/tigergraph/ecosys/tree/benchmark/benchmark}}.

GSQL represents a sweet spot in the trade-off between abstraction level and expressivity:
it is sufficiently high-level to allow declarative SQL-style programming, yet
sufficiently expressive to specify sophisticated iterative graph algorithms and configurable
\darpe\ semantics. These are traditionally coded in general-purpose languages like C++ and Java
and available only as built-in library functions in other graph query languages such as
Gremlin and Cypher, with the drawback that advanced programming expertise is required for
customization.

The GSQL query language shows that the choice
between declarative SQL-style and NoSQL-style programming over graph data is a false choice,
as the two are eminently compatible. GSQL also shows a way to unify the querying of
relational tabular and graph data.

GSQL is still evolving, in response to our experience with customer deployment.
We are also responding to the experiences of the graph developer community at large,
as TigerGraph is a participant in current ANSI standardization working groups for graph query languages
and graph query extensions for SQL.

\bibliographystyle{ACM-Reference-Format}
\bibliography{main}

\appendix
\section{GSQL’s Main Built-In Accumulator Types}
\label{sec:built-in-accums}

GSQL comes with a list of pre-defined accumulators, some of which we
detail here. For details on GSQL’s accumulators and more supported types, see the online
documentation at
\myhref{https://docs.tigergraph.com/dev/gsql-ref}{https://docs.tigergraph.com/dev/gsql-ref}).

{\tt SumAccum<N>},
where N is a numeric type. This accumulator holds an internal value of 
type N, accepts inputs of type N and aggregates them into the internal
value using addition.\\

{\tt MinAccum<O>},
where {\tt O} is an ordered type. It computes the
minimum value of its inputs of type {\tt O}.\\

{\tt MaxAccum<O>},
as above, swapping max for min aggregation.\\

{\tt AvgAccum<N>},
where {\tt N} is a numeric type. This accumulator
computes the average of its inputs of type {\tt N}.
It is implemented in an order-invariant way by internally
maintaining both the sum and the count of the inputs seen so far.\\

{\tt OrAccum},
which aggregates its boolean inputs using logical disjunction.\\

{\tt AndAccum},
which aggregates its boolean inputs using logical conjunction.\\

{\tt MapAccum<K,V>}
stores an internal value of map type, where {\tt K} is the type of keys and
{\tt V} the type of values.
{\tt V} can itself be an accumulator type, thus specifying how to aggregate values mapped
to the same key.\\

{\tt HeapAccum<T>(capacity, field\_1 [ASC|DESC], field\_2 [ASC|DESC], …, field\_n [ASC|DESC])}
implements a priority queue where 	
{\tt T} is a tuple type whose fields include {\tt field\_1} through {\tt field\_n},
each of ordered type,	
{\tt capacity} is the integer size of the priority queue, and the remaining arguments
specify a lexicographic order for sorting the tuples in the priority queue
(each field may be used in ascending or descending order).

\section{Benchmark}
\label{sec:benchmark}

To evaluate TigerGraph experimentally, we designed a benchmark for MPP graph databases
that allows us to explore the mult-cores/single-machine setting as well as the scale-out
performance for computing and storage obtained from a cluster of machines.

In this section, we examine the data loading, query performance, and
their associated resource usage for TigerGraph.
For results on running the same experiments for other graph database systems,
see~\cite{TigerGraph:benchmark}.

The experiments test the performance of data loading and querying.


\myparagraph{Data Loading} To study loading performance, we measure
\begin{itemize}
\item Loading time;
\item Storage size of loaded data; and
\item Peak CPU and memory usage on data loading.
\end{itemize}
The data loading test will help us understand the loading speed, TigerGraph's
data storage compression effect, and the loading resource requirements.

\myparagraph{Querying} We study performance for the following representative queries
over the schema available \myhref{https://github.com/tigergraph/ecosys/blob/benchmark/benchmark/tigergraph/graph500_setup.gsql}{on GitHub}.
\begin{itemize}
\item Q1. K-hop neighbor count (code available \myhref{https://github.com/tigergraph/ecosys/blob/benchmark/benchmark/tigergraph/khop.gsql}{here}). We measure the query response time and throughput.
\item Q2. Weakly-Connected Components (\myhref{https://github.com/tigergraph/ecosys/blob/benchmark/benchmark/tigergraph/wcc-graph500.gsql}{here}). We measure the query response time.
\item Q3. PageRank (\myhref{https://github.com/tigergraph/ecosys/blob/benchmark/benchmark/tigergraph/pg-graph500.gsql}{here}). We measure the query response time for 10 iterations. 
\item Peak CPU and memory usage for each of the above query. 
\end{itemize}
The query test will reveal the performance of the interactive query
workload (Q1) and the analytic query workload (Q2 and Q3). All queries are first benchmarked on a
single EC2 machine, followed by Q1 employed to test TigerGraph's scale-up capability on EC2 R4
family instances, and Q3 used to test the scale-out capability on clusters of various sizes. 

With these tests, we were able to show the following properties of TigerGraph.
\begin{itemize}
\item High performance of loading: >100GB/machine/hour.
\item Deep-link traversal capability: >10 hops traversal on a billion-edge
scale real social graph. 
\item Linear scale-up query performance: the query throughput linearly
increases with the number of cores on a single machine.
\item Linear scale-out query performance: the analytic query response
time linearly decreases with the number of machines. 
\item Low and constant memory consumption and consistently high CPU utilization.
\end{itemize}

\subsection{Experimental Setup}

\begin{table}
\centering
\caption{Datasets}
\label{table:dataset}
\begin{tabular}{|l|p{3.2cm}|l|l|}
\hline 
\textbf{Name} & \textbf{Description}  & \textbf{Vertices}  & \textbf{Edges}\\
\hline 
graph500 & Synthetic Kronecker graph\footnotemark[1]& 2.4M & 67M\tabularnewline
\hline 
twitter & Twitter user-follower directed graph\footnotemark[2] & 41.6M & 1.47B\tabularnewline
\hline 
\end{tabular}
\end{table}
\footnotetext[1]{http://graph500.org}
\footnotetext[2]{http://an.kaist.ac.kr/traces/WWW2010.html}

\begin{table}
\centering
\caption{Cloud hardware and OS}
\label{table:hardware}
\begin{tabular}{|l|p{0.9cm}|p{1cm}|p{1.5cm}|l|}
\hline 
\textbf{EC2} & \small{\textbf{vCPUs}} & \small{\textbf{Mem}}  & \textbf{SSD(GP2)} & \textbf{OS}\tabularnewline
\hline 
r4.2xlarge & 8 & 61G & 200G & \small{Ubuntu 14.04}\tabularnewline
\hline 
r4.4xlarge & 16 & 122G & 200G & \small{Ubuntu 14.04}\tabularnewline
\hline 
r4.8xlarge & 32 & 244G & 200G & \small{Ubuntu 14.04}\tabularnewline
\hline 
\end{tabular}
\end{table}

\noindent\textbf{Datasets.} The experiment uses the two data sets described in
Table \ref{table:dataset}. One synthetic and one real. For each graph, the raw data are formatted as a
single tab-separated edge list. Each row contains two columns, representing the source and the target
vertex id, respectively. Vertices do not have attributes, so there is no need for a separate vertex list. 

\vspace{1ex}\noindent\textbf{Software. }For the single-machine experiment, we used the freely
available TigerGraph Developer Edition 2.1.4. For the multi-machine experiment, we used TigerGraph
Enterprise Edition 2.1.6. All queries are written in GSQL. The graph database engine
is written in C++ and can be run on Linux/Unix or container environments.
For resource profiling, we used Tsar\footnote{https://github.com/alibaba/tsar}. 

\vspace{1ex}\noindent\textbf{Hardware.} We ran the single-machine experiment on an Amazon EC2
r4.8xlarge instance type. For the single machine scale-up experiment, we used
r4.2xlarge, r4.4xlarge and r4.8xlarge instances.
The multi-machine experiment used r4.2xlarge instances to form different-sized clusters.

\subsection{Data Loading Test}
\vspace{1ex}\noindent\textbf{Methodology.} For both data sets, we used GSQL DDL to create a graph
schema containing one vertex type and one edge type. The edge type is a directed edge connecting
the vertex type to itself. A declarative loading job was written in the GSQL loading language.
There is no pre-(e.g. extracting unique vertex ids) or post-processing (e.g. index building). 
\begin{table}
\centering
\caption{Loading Results}
\label{table:loading}
\begin{tabular}{|l|p{1.5cm}|l|l|}
\hline 
\textbf{Name} & \textbf{Raw Size}  & \textbf{TigerGraph Size}  & \textbf{Duration}\\
\hline 
graph500 & 967M & 482M & 56s\tabularnewline
\hline 
twitter & 24,375M & 9,500M & 813s\tabularnewline
\hline 
\end{tabular}
\end{table}

\vspace{1ex}\noindent\textbf{Results Discussion.} TigerGraph loaded the twitter data at the speed of 100G/Hour/Machine (Table \ref{table:loading}). TigerGraph automatically encodes and compresses the raw data to less than half its original size--Twitter (2.57X compression), Graph500(2X compression). The size of the loaded data is an important consideration for system cost and performance. All else being equal, a compactly stored database can store more data on a given machine and has faster access times because it gets more cache and memory page hits. 
The measured peak memory usage for loading was 2.5\% for graph500 and 7.08\% for Twitter,
while CPU peak usage was 48.4\% for graph500 and 60.3\% for Twitter.

\subsection{Querying Test}

\subsubsection{Q1}\ \\

\myparagraph{A. Single-machine k-hop test}
The k-hop-path neighbor count query, which asks for the total count of the vertices which have a
k-length simple path from a starting vertex is a good stress test for  graph traversal performance.

\myparagraph{Methodology} For each data set, we count all k-hop-path endpoint vertices for 300 fixed random seeds \emph{sequentially}. By ``fixed random seed", we mean we make a one-time random selection of N vertices from the graph and save this list as repeatable input condition for our tests. We measure the average query response time for k=1,2, 3,6,9,12 respectively.

\myparagraph{Results Discussion} For both data sets, TigerGaph can answer deep-link k-hop queries as shown in Tables \ref{table:g500_k} and \ref{table:twitter_k}. For graph500, it can answer 12-hop queries in circa 4 seconds. For the billion-edge Twitter graph, it can answer 12-hop queries in under 3 minutes. We note the significance of this accomplishment by pointing out that starting from 6 hops and above, the average number of k-hop-path neighbors per query is around 35 million. In contrast, we have benchmarked 5 other top commercial-grade property graph databases on the same tests, and all of them started failing on some 3-hop queries, while failing on all 6-hop queries \cite{TigerGraph:benchmark}.  Also, note that the peak memory usage stays constant regardless of k. The constant memory footprint enables TigerGraphto follow links to  \emph{unlimited} depth to perform what we term ``deep-link analytics''. CPU utilization increases with the hop count. This is expected as each hop can discover new neighbors at an exponential growth rate, thus leading to more CPU usage.

\begin{table}
\centering
\caption{Graph500 - K-hop Query on r4.8xlarge }
\label{table:g500_k}
\begin{tabular}{|l|p{1.8cm}|p{1.5cm}|p{1cm}|l|}
\hline 
\textbf{K-hop} & \small{\textbf{\small{Avg RESP}}} & \small{\textbf{Avg N}}  & \textbf{CPU} & \textbf{MEM}\tabularnewline
\hline 
1 & 5.95ms & 5128 & 4.3\% & 3.8\%\tabularnewline
\hline 
2 & 69.94ms & 488,723 & 72.3\% & 3.8\%\tabularnewline
\hline 
3 & 409.89ms & 1,358,948 & 84.7\% & 3.7\%\tabularnewline
\hline 
6 & 1.69s & 1,524,521 & 88.1\% & 3.7\%\tabularnewline
\hline 
9 & 2.98s & 1,524,972 & 87.9\% & 3.7\%\tabularnewline
\hline 
12 & 4.25s & 1,524,300 & 89.0\% & 3.7\%\tabularnewline
\hline 
\end{tabular}
\end{table}

\begin{table}
\centering
\caption{Twitter - K-hop Query on r4.8xlarge Instance}
\label{table:twitter_k}
\begin{tabular}{|l|p{1.8cm}|p{1.5cm}|p{0.8cm}|l|}
\hline 
\textbf{K-hop} & \small{\textbf{\small{Avg RESP}}} & \small{\textbf{Avg N}}  & \textbf{CPU} & \textbf{MEM}\tabularnewline
\hline 
1 & 22.60ms & 106,362 & 9.8\% & 5.6\%\tabularnewline
\hline 
2 & 442.70ms & 3,245,538 & 78.6\% & 5.7\%\tabularnewline
\hline 
3 & 6.64s & 18,835,570 & 99.9\% & 5.8\%\tabularnewline
\hline 
6 & 56.12s & 34,949,492 & 100\% & 7.6\%\tabularnewline
\hline 
9 & 109.34s & 35,016,028 & 100\% & 7.7\%\tabularnewline
\hline 
12 & 163.00s & 35,016,133 & 100\% & 7.7\%\tabularnewline
\hline 
\end{tabular}
\end{table}

\vspace{1ex}\noindent\textbf{B. Single machine k-hop scale-up test}
In this test, we study TigerGraph's scale-up ability,
i.e. the ability to increase performance with increasing number of cores on a single machine.

\vspace{1ex}\noindent\textbf{Methodology.} 
The same  query workload was tested on machines with different number of cores:
for a given k and a data set, 22 client threads keep sending the 300 k-hop queries
to a TigerGraph server concurrently. When a query responds, the client thread responsible
for that query will move on to send the next unprocessed query.
We tested the same k-hop query workloads on three different machines on the R4
family: r4.2xlarge (8vCPUs), r4.4xlarge (16vCPUs), and r4.8xlarge (32vCPUs).

\begin{figure}
    \includegraphics[width=1\columnwidth]{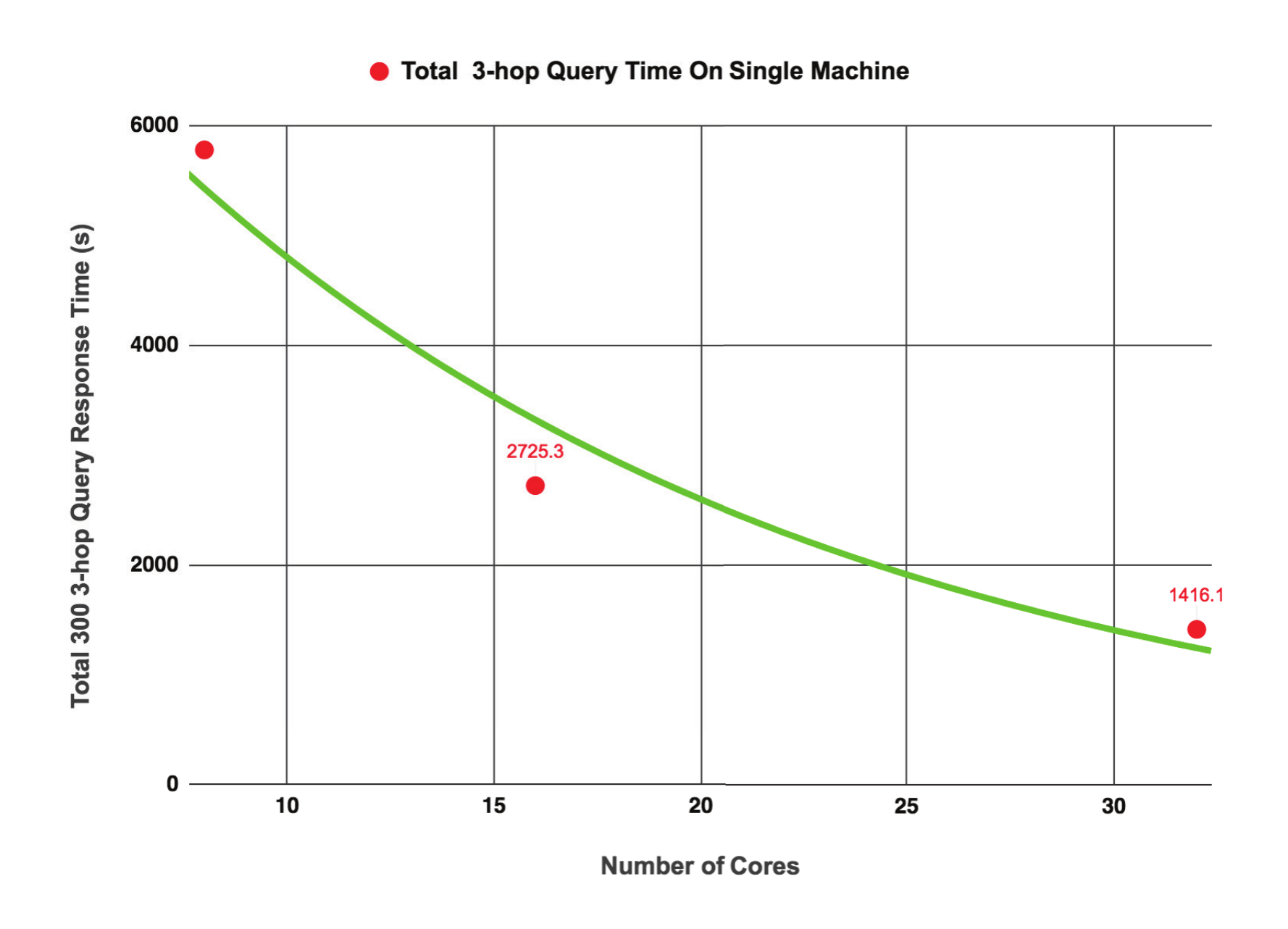}
    \caption{Scale Up 3-hop Query Throughput On Twitter Data By Adding More Cores.}
    \label{fig:scaleup}
\end{figure}

\vspace{1ex}\noindent\textbf{Results Discussion.} As shown by Figure \ref{fig:scaleup}, TigerGraph linearly scales up the 3-hop query throughput test with the number of cores: 300 queries on the Twitter data set take 5774.3s on a 8-core machine, 2725.3s on a 16-core machine, and 1416.1s on a 32-core machine. These machines belong to the same EC2 R4 instance family, all having the same memory bandwidth- DDR4 memory, 32K L1, 256K L2, and 46M L3 caches, the same CPU model- dual socket Intel Xeon E5 Broadwell processors (2.3 GHz).

\begin{table}[]
\caption{Analytic Queries }
\label{table:analytic}
\begin{tabular}{|l|l|l|l|l|l|l|}
\hline
\multicolumn{1}{|c|}{\multirow{2}{*}{\textbf{Data}}} & \multicolumn{3}{c|}{\textbf{WCC}}                                                                          & \multicolumn{3}{c|}{\textbf{10-iter PageRank}}                                                             \\ \cline{2-7} 
\multicolumn{1}{|c|}{}                               & \multicolumn{1}{c|}{\textbf{Time}} & \multicolumn{1}{c|}{\textbf{CPU}} & \multicolumn{1}{c|}{\textbf{Mem}} & \multicolumn{1}{c|}{\textbf{Time}} & \multicolumn{1}{c|}{\textbf{CPU}} & \multicolumn{1}{c|}{\textbf{Mem}} \\ \hline
G500                                                 & 3.1s                               & 81.0\%                            & 2.2\%                             & 12.5s                              & 89.5\%                            & 2.3\%                             \\ \hline
Twitter                                              & 74.1s                              & 100\%                             & 7.7\%                             & 265.4s                             & 100\%                             & 5.6\%                             \\ \hline
\end{tabular}
\end{table}

\subsubsection{Q2 and Q3}\ \\
\label{sec:benchmark-full-graph}

{\em Full-graph queries}
examine the entire graph and compute results which describe the characteristics of the graph.

\myparagraph{Methodology}
We select two full-graph queries, namely weakly connected component labeling and PageRank. A weakly connected component (WCC) is the maximal set of vertices and their connecting edges which can reach one another, if the direction of directed edges is ignored. The WCC query finds and labels all the WCCs in a graph. This query requires that every vertex and every edge be traversed. PageRank is an iterative algorithm which traverses every edge during every iteration and computes a score for each vertex. After several iterations, the scores will converge to steady state values. For our experiment, we run 10 iterations. Both algorithms are implemented in the GSQL language.

\vspace{1ex}\noindent\textbf{Results Discussion.} TigerGraph performs analytic queries very fast with excellent memory footprint. Based on Table \ref{table:analytic}, for the Twitter data set, the storage size is 9.5G and the peak WCC memory consumption is 244G*7.7\%, which is 18G and PageRank peak memory consumption is 244G*5.6\%, which is 13.6G. Most of the time, the 32 cores' total utilization is above 80\%, showing that TigerGraph exploits multi-core capabilities efficiently.

\subsubsection{Q3 Scale Out}\ \\
\label{sec:benchmark-scale-out}

\begin{figure}
    \includegraphics[width=1\columnwidth]{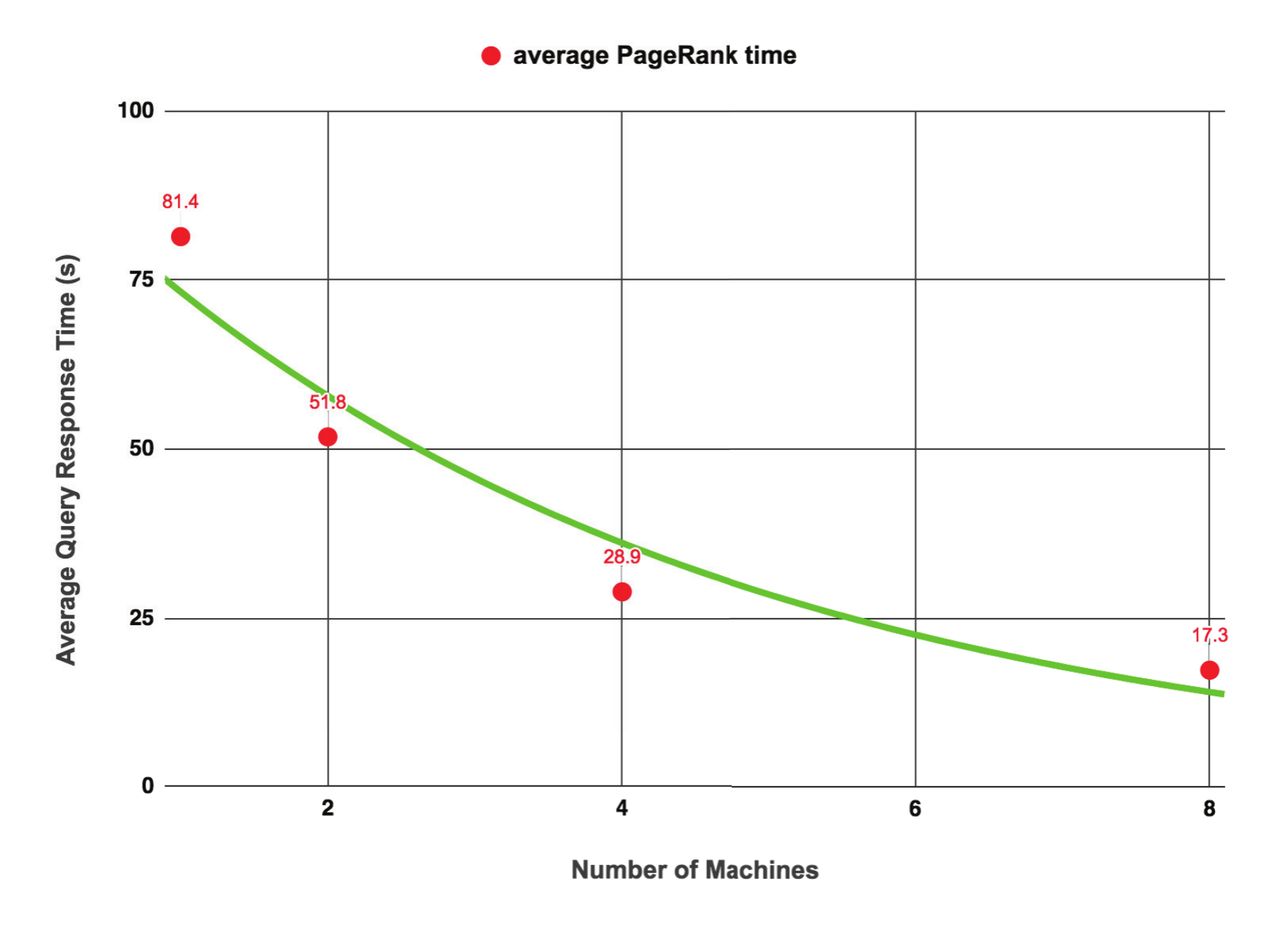}
    \caption{Scale Out 10-iteration PageRank Query Response Time On Twitter Data By Adding More Machines.}
    \label{fig:scaleout}
\end{figure}

 All previous tests ran in a single-machine setting.
 This test looks at how TigerGraph's performance scales as the number of compute servers increases.
 
\myparagraph{Methodology} 
For this test, we used a more economical Amazon EC2 instance type (r4.2xlarge: 8 vCPUs, 61GiB memory, and 200GB attached GP2 SSD). When the Twitter dataset (compressed to 9.5 GB by TigerGraph) is distributed across 1 to 8 machines, 60GiB memory and 8 cores per machine is more than enough. For larger graphs or for higher query throughput, more cores may help; TigerGraph provides settings to tune memory and compute resource utilization. Also, to run on a cluster, we switched from the TigerGraph Developer Edition (v2.1.4) to the Enterprise Edition (v2.1.6).
We used the Twitter dataset and ran the PageRank query for 10 iterations. We repeated this three times and averaged the query times. We repeated the tests for clusters containing 1, 2, 4 and 8 machines. For each cluster, the Twitter graph was partitioned into equally-sized segments across all the machines being used.

\myparagraph{Results Discussion}
PageRank is an iterative algorithm which traverses every edge during every iteration. This means there is much communication between the machines, with information being sent from one partition to another. Despite this communication overhead, TigerGraph's Native MPP Graph database architecture still succeeds in achieving a 6.7x speedup with 8 machines (Figure \ref{fig:scaleout}), showing good linear scale-out performance.

\subsection{Reproducibility} All of the files needed to reproduce the  tests (datasets, queries, scripts, input parameters, result files, and general instructions) are available on GitHub\cite{TigerGraph:benchmark}.

\lstdefinestyle{Grammar}{
  language=SQL,
  otherkeywords={HAVING, QUERY, FOR, for, foreach, GRAPH, WITH, BEGIN, END, ACCUM, POST, WHILE, while, break, continue, DO, IF, if,  VERTEX, vertex, RETURN, TO, UNDIRECTED, DIRECTED, UN, EDGE, REVERSE, DISCRIMINATOR, STRING, BETWEEN, IN, LIKE, IS, is, INTERSECT, intersect, TERSECT, tersect, DO, do, SetAccum, BagAccum, HeapAccum, OrAccum, AndAccum, MaxAccum, MinAccum, SumAccum, AvgAccum, ListAccum, ArrayAccum, StringAccum, MapAccum, GroupByAccum, BitwiseOrAccum, BitwiseAndAccum, ASC, DESC, +=, =, minus, -, ., :, ;, _, <, >, @, ', Id, NumConst, StringConst, DateTimeConst, true, false, type, int, uint, long, float, double, string, boolean, datetime, Tuple, set, bag, map, range},
  numbers=left,
  stepnumber=0,
  numbersep=10pt,
  tabsize=4,
  showspaces=false,
  showstringspaces=false
}

\lstset{
    escapeinside={(*@}{@*)},
}

\section{Formal Syntax}

In the following, terminal symbols are shown in bold font and not further defined
when their meaning is self-understood.

\subsection{Declarations}

\lstset{style=Grammar}
\begin{lstlisting}
decl (*@$\rightarrow$@*) accType gAccName (= expr)?  (*@{\mybf{;}@*)
     | accType vAccName  (= expr)?  (*@{\mybf{;}@*)
     | baseType var (= expr)?  (*@{\mybf{;}@*)

gAccName (*@$\rightarrow$@*) @@Id
vAccName (*@$\rightarrow$@*) @Id


accType (*@$\rightarrow$@*) SetAccum   <baseType>
        |  BagAccum   <baseType>
        |  HeapAccum<tupleType>(*@{\mybf{(}@*)capacity ((*@{\mybf{,}@*) Id dir?)+(*@{\mybf{)}@*)
        |  OrAccum
        |  AndAccum
        |  (*@{\tt\textbf{BitwiseOrAccum}@*)
        |  (*@{\tt\textbf{BitwiseAndAccum}@*)
        |  MaxAccum   <orderedType>
        |  MinAccum   <orderedType>
        |  SumAccum   <numType|string>
        |  AvgAccum   <numType>
        |  (*@{\tt\textbf{ListAccum}@*)  <(*@{\tt type}@*)>
        |  ArrayAccum <(*@{\tt type}@*)> dimension+
        |  MapAccum   <baseType (*@{\mybf{,}@*) (*@{\tt type}@*)>
        |  GroupByAccum <baseType Id
                         ((*@{\mybf{,}@*) baseType Id)* (*@{\mybf{,}@*)
                         accType>

baseType (*@$\rightarrow$@*) orderedType
         |  boolean
         |  datetime
         |  (*@{\tt\textbf{edge}@*) (< edgeType >)?
         |  tupleType

orderedType (*@$\rightarrow$@*) numType
            | string
            | vertex (< (*@{\tt vertexType}@*) >)?
       
numType (*@$\rightarrow$@*) int
        |  (*@{\tt\textbf{uint}@*)
        |  float
        |  double

capacity (*@$\rightarrow$@*) NumConst | paramName

paramName (*@$\rightarrow$@*) Id

tupleType (*@$\rightarrow$@*) Tuple < baseType Id ((*@{\mybf{,}@*) baseType Id)* >

dir (*@$\rightarrow$@*) ASC | DESC

dimension (*@$\rightarrow$@*) (*@{\mybf{[}@*)expr?(*@{\mybf{]}@*)

(*@type\ $\rightarrow$@*) baseType | accType         
\end{lstlisting}

\subsection{\darpe s}
\begin{lstlisting}
darpe (*@$\rightarrow$@*) edgeType
      | edgeType>
      | <edgeType
      | darpe(*@{\mybf{*}}@*) bounds?
      | (*@{\mybf{(}}@*)darpe(*@{$\pmb{\bm{)}}$}@*)
      | darpe ((*@{\mybf{.}@*) darpe)+
      | darpe ((*@{\mybf{|}}@*) darpe)+

edgeType (*@$\rightarrow$@*) Id
      
bounds (*@$\rightarrow$@*) NumConst (*@{\mybf{..}@*) NumConst
       | (*@{\mybf{..}@*) NumConst
       | NumConst (*@{\mybf{..}@*)
       | NumConst       
\end{lstlisting}

\subsection{Patterns}

\begin{lstlisting}
  pathPattern  (*@$\rightarrow$@*)
     vTest ((*@{\mybf{:}@*) var)?
  |  pathPattern
       -(*@{\mybf{(}}@*)darpe ((*@{\mybf{:}@*) var)?(*@{\mybf{)}}@*)- vTest ((*@{\mybf{:}@*) var)?

  pattern (*@$\rightarrow$@*) pathPattern ((*@{\mybf{,}@*) pathPattern)*
           
  vTest  (*@$\rightarrow$@*) (*@{\mybf{\_}@*)
         | (*@{\tt vertexType}@*) ((*@{\mybf{|}}@*) (*@{\tt vertexType}@*))*

  var (*@$\rightarrow$@*) Id

  (*@{\tt vertexType}@*) (*@$\rightarrow$@*) Id
\end{lstlisting}

\subsection{Atoms}
\begin{lstlisting}
atom (*@$\rightarrow$@*) relAtom | graphAtom

relAtom   (*@$\rightarrow$@*) tableName AS? var
tableName (*@$\rightarrow$@*) Id

graphAtom  (*@$\rightarrow$@*) (graphName AS?)? pattern
graphName  (*@$\rightarrow$@*) Id
\end{lstlisting}

\subsection{\from\ Clause}
\begin{lstlisting}
fromClause (*@$\rightarrow$@*) FROM atom ((*@{\mybf{,}@*) atom)*
\end{lstlisting}

\subsection{Terms}

\begin{lstlisting}
term (*@$\rightarrow$@*) constant
     |  var
     |  var(*@{\mybf{.}@*)attribName
     |  gAccName
     |  gAccName(*@{\mybf{'}@*)
     |  var(*@{\mybf{.}@*)vAccName
     |  var(*@{\mybf{.}@*)vAccName(*@{\mybf{'}@*)
     |  var(*@{\mybf{.}@*)type
     
constant (*@$\rightarrow$@*) NumConst
         | StringConst
         | DateTimeConst
         | true
         | false
         
attribName (*@$\rightarrow$@*) Id
\end{lstlisting}
     
\subsection{Expressions}
\begin{lstlisting}
  expr (*@$\rightarrow$@*) term
       | (*@{\mybf{(}}@*) expr (*@{\mybf{)}}@*)
       | (*@{\mybf{-}@*) expr
       | expr arithmOp  expr
       | not expr
       | expr logicalOp expr
       | expr (*@{\tt setOp}@*) expr
       | expr between expr and expr
       | expr not? in expr
       | expr like expr
       | expr is not? null
       | fnName (*@{\mybf{(}}@*) exprs? (*@{\mybf{)}}@*)
       | case
           (when condition then expr)+
           (else expr)?
         end
       | case expr
           (when constant then expr)+
           (else expr)?
         end
       | (*@{\mybf{(}}@*)exprs?(*@{\mybf{)}}@*)
       | (*@{\mybf{[}}@*)exprs?(*@{\mybf{]}}@*)
       | (*@{\mybf{(}}@*) exprs -> exprs (*@{\mybf{)}}@*) // (*@{\tt MapAccum input}@*)
       | expr arrayIndex+  // (*@{\tt ArrayAccum access}@*)

arithmOp (*@$\rightarrow$@*) (*@$\mybf{*}\ |\ \mybf{/}\ |\ \mybf{\%}\ |\ \mybf{+}\ |\ \mybf{-}\ |\ \mybf{\&}\ |\ \mybf{|}$@*)

logicalOp (*@$\rightarrow$@*) and | or

(*@{\tt setOp}@*) (*@$\rightarrow$\ {\tt\textbf{intersect}}@*) | union | minus  

exprs (*@$\rightarrow$@*) expr((*@{\mybf{,}@*) exprs)*

arrayIndex  (*@$\rightarrow$@*) (*@{\mybf[}@*)expr(*@{\mybf]}@*)
\end{lstlisting}

\subsection{\where\ Clause}
\begin{lstlisting}
  whereClause (*@$\rightarrow$@*) WHERE condition

  condition (*@$\rightarrow$@*) expr
\end{lstlisting}

\subsection{\accum\ Clause}
\begin{lstlisting}
  accClause (*@$\rightarrow$@*) ACCUM stmts
  
  stmts (*@$\rightarrow$@*) stmt ((*@{\mybf{,}@*) stmt)*
  
  stmt (*@$\rightarrow$@*) varAssignStmt
       |  vAccUpdateStmt
       |  gAccUpdateStmt
       |  (*@{\tt forStmt}@*)
       |  caseStmt
       |  (*@{\tt ifStmt}@*)
       |  (*@{\tt whileStmt}@*)
       
  varAssignStmt (*@$\rightarrow$@*) baseType? var = expr

  vAccUpdateStmt (*@$\rightarrow$@*) var(*@{\mybf{.}@*)vAccName = expr
                 |  var(*@{\mybf{.}@*)vAccName (*@{\tt\textbf{+=}}@*) expr

  gAccUpdateStmt (*@$\rightarrow$@*) gAccName = expr
                 |  gAccName (*@{\tt\textbf{+=}}@*) expr

  (*@{\tt forStmt}@*) (*@$\rightarrow$@*) foreach var in expr do stmts end
          |  foreach (var ((*@{\mybf{,}@*) var)*) in expr do stmts end
          |  foreach var in range (*@{\mybf{(}}@*)expr (*@{\mybf{,}}@*) expr(*@{\mybf{)}}@*) do stmts end
          
  caseStmt (*@$\rightarrow$@*) case
                (when condition then stmts)+
                (else stmts)?
              end
           |  case expr
                (when constant then stmts)+
                (else stmts)?
              end

  (*@{\tt ifStmt}@*) (*@$\rightarrow$@*) if condition then stmts (else stmts)? end

  (*@{\tt whileStmt}@*) (*@$\rightarrow$@*) while condition limit expr do body end
  body      (*@$\rightarrow$@*) bodyStmt ((*@{\mybf{,}@*) bodyStmt)*
  bodyStmt  (*@$\rightarrow$@*) stmt
            |  continue
            |  break
             
\end{lstlisting}

\subsection{\paccum\ Clause}
\begin{lstlisting}
  pAccClause (*@$\rightarrow$@*) POST_ACCUM stmts
\end{lstlisting}

\subsection{\select\ Clause}
\begin{lstlisting}
  selectClause (*@$\rightarrow$@*) SELECT outTable ((*@{\mybf{;}@*) outTable)*

  outTable (*@$\rightarrow$\ {\tt\textbf{DISTINCT}}@*)? col ((*@{\mybf{,}@*) col)* INTO tableName

  col (*@$\rightarrow$@*) expr (AS colName)?

  tableName (*@$\rightarrow$@*) Id

  colName (*@$\rightarrow$@*) Id
\end{lstlisting}

\subsection{GROUP BY Clause}
\begin{lstlisting}
  groupByClause (*@$\rightarrow$@*) GROUP BY exprs ((*@{\mybf{;}@*) exprs)*
\end{lstlisting}

\subsection{HAVING Clause}
\begin{lstlisting}
  havingClause (*@$\rightarrow$ {\tt\textbf{HAVING}}@*) condition ((*@{\mybf{;}@*) condition)*
\end{lstlisting}

\subsection{ORDER BY Clause}
\begin{lstlisting}
  orderByClause (*@$\rightarrow$@*) ORDER BY oExprs ((*@{\mybf{;}@*) oExprs)*

  oExprs (*@$\rightarrow$@*) oExpr((*@{\mybf{,}@*) oExpr)*

  oExpr (*@$\rightarrow$@*) expr dir?
\end{lstlisting}

\subsection{LIMIT Clause}
\begin{lstlisting}
  limitClause (*@$\rightarrow$@*) LIMIT expr ((*@{\mybf{;}@*) expr)*
\end{lstlisting}

\subsection{Query Block Statements}
\begin{lstlisting}
  queryBlock (*@$\rightarrow$@*) selectClause
             |  fromClause
             |  whereClause?
             |  accClause?
             |  pAccClause?
             |  groupByClause?
             |  havingClause?
             |  orderByClause?
             |  limitClause?
\end{lstlisting}

\pagebreak
\subsection{Query}
\begin{lstlisting}
  query (*@$\rightarrow$@*) CREATE QUERY Id (*@(@*)params?(*@)@*)
           (FOR GRAPH graphName)? (*@${\mybf\{}$@*)
             decl*
             qStmt*
             (RETURN expr)?
           (*@${\mybf\}}$@*)

  params (*@$\rightarrow$@*) param ((*@{\mybf{,}@*) param)*

  param (*@$\rightarrow$@*) paramType paramName

  paramType (*@$\rightarrow$@*) baseType
            |  set<baseType>
            |  bag<baseType>
            |  map<baseType (*@{\mybf{,}@*) baseType>

  qStmt (*@$\rightarrow$@*) stmt (*@{\mybf{;}@*) | queryBlock (*@{\mybf{;}@*)

\end{lstlisting}

\section{GSQL Formal Semantics}
GSQL expresses queries over standard SQL tables and over
graphs in which both directed and undirected edges may coexist,
and whose vertices and edges carry data (attribute name-value maps).

The core of a GSQL query is the SELECT-FROM-WHERE block modeled after SQL,
with the FROM clause specifying a pattern to be matched against the graphs and the tables.
The pattern contains vertex, edge and tuple variables and each match induces a
variable binding for them. The WHERE and SELECT clauses treat these variables as in SQL.
GSQL supports an additional \accum\ clause that is used to update accumulators.

\subsection{The Graph Data Model}
In TigerGraph's data model, graphs allow both directed and undirected edges.
Both vertices and edges can carry data (in the form of attribute name-value maps).
Both vertices and edges are typed.
  
Let
$\vIds$ denote a countable set of vertex ids,
$\eIds$ a countable set of edge ids disjoint from $\vIds$,
$\attrNames$ a countable set of attribute names,
$\vTyNames$ a countable set of vertex type names,
$\eTyNames$ a countable set of edge type names.

Let $\calD$ denote an infinite domain (set of values),
which comprises
\begin{itemize}
\item
  all numeric values,
\item
  all string values,
\item
  the boolean constants {\tt{\textbf{true}}} and {\tt{\textbf{false}}},
\item
  all datetime values,
\item
  $\vIds$,
\item
  $\eIds$,
\item
  all sets of values (their sub-class is denoted $\{\calD\}$),
\item
  all bags of values ($\bag{D}$),
\item
  all lists of values ($[\calD]$), and
\item
  all maps (sets of key-value pairs, $\{ \calD \mapsto \calD \}$).
\end{itemize}

A graph is a tuple
$$
G = (V, E, \epts, \tyV, \tyE, \data)
$$

where
\begin{itemize}
\item
  $V \subset \vIds$ is a finite set of vertices.
\item
  $E \subset \eIds$  is a finite set of edges.
\item
$
\epts: E \rightarrow 2^{V \times\ V}
$
is a function that associates with an edge $e$ its endpoint vertices.
If $e$ is directed, $\epts(e)$
is a singleton set containing a (source,target) vertex pair.
If $e$ is undirected, $\epts(e)$ is a
set of two pairs, corresponding to both possible orderings of the endpoints.
\item
  $\tyV: V \rightarrow \vTyNames$ is a function that associates a type name to each vertex.
\item
  $\tyE: E \rightarrow \eTyNames$ is a function that associates a type name to each edge.
\item
  $\data: (V \cup E) \times \attrNames \rightarrow \calD$ is a function that associates domain values to
  vertex/edge attributes (identified by the vertex/edge id and the attribute name).
\end{itemize}

\subsection{Contexts}
GSQL queries are composed of multiple statements.
A statement may refer to intermediate results provided
by preceding statements (e.g. global variables, temporary tables, accumulators).
In addition, since GSQL queries can be parameterized
just like SQL views, statements may refer to
parameters whose value is provided by the initial query call.

A statement must therefore be evaluated in a {\em context} which provides the values for
the names referenced by the statement.
We model a context as a map from the names to the values of
parameters/global variables/temporary tables/accumulators, etc.

Given a context map $\ctx$ and a name $n$, $\ctx(n)$
denotes the value associated to $n$ in $\ctx$~\footnote{
  In the remainder of the presentation we assume that the query has passed all
  appropriate semantic and type checks and therefore $\ctx(n)$ is defined for every $n$
  we use.}

$\dom(\ctx)$ denotes the {\em domain} of context \ctx, i.e. the set of names
which have an associated value in \ctx (we say that these names are {\em defined} in \ctx).

\paragraph{Overriding Context Extension} When a new variable is introduced by a statement
operating within a context $\ctx$, the context needs to be extended with the new variable's
name and value.

$$
\override{\ctx}{\{n \mapsto v\}}
$$
denotes a new context obtained
by modifying a {\em copy} of $\ctx$ to associate name $n$ with value $v$
(overwriting any pre-existing entry for $n$).

Given contexts
$\ctx_1$ and
$\ctx_2 = \{n_i \mapsto v_i \}_{1 \leq i \leq k}$, we say that $\ctx_2$ {\em overrides}
$\ctx_1$, denoted $\override{\ctx_1}{\ctx_2}$ and defined as:
\begin{eqnarray*}
\override{\ctx_1}{\ctx_2} & = & c_k \mathit{\ where}\\
c_0 & = & \ctx_1, \mathit{\ and}\\
c_i & = & \override{c_{i-1}}{\{n_i \mapsto v_i\}}, \mathit{\ for}\ 1 \leq i \leq k
\end{eqnarray*}

\paragraph{Consistent Contexts}
We call two contexts $\ctx_1,\ctx_2$ {\em consistent} if they agree on
every name in the intersection of their domains. That is, for
each $x \in \dom(\ctx_1) \cap \dom(\ctx_2)$, $\ctx_1(x) = \ctx_2(x)$.

\paragraph{Merged Contexts}
For consistent contexts, we can define $\ctx_1 \cup \ctx_2$, which denotes
the {\em merged context} over the union of the domains of $\ctx_1$ and $\ctx_2$:
$$
\ctx_1 \cup \ctx_2 (n) =
\left\{
\begin{array}{l@{, \ \ }l}
  \ctx_1(n) & \mbox{if\ } n \in \dom(ctx_1)\\
  \ctx_2(n) & \mbox{otherwise}
\end{array}
\right.
$$

\subsection{Accumulators}
A GSQL query can declare accumulators whose names come from 
\gAccNames, a countable set of global accumulator names and from
\vAccNames, a disjoint countable set of vertex accumulator names.

Accumulators are data types that store an internal value and take
inputs that are aggregated into this internal value using a binary operation.
GSQL distinguishes among two accumulator flavors:
\begin{itemize}
\item
  {\em Vertex accumulators} are attached to vertices,
  with each vertex storing its own local accumulator instance.
\item
 {\em Global accumulators} have a single instance.
\end{itemize}
Accumulators are polymorphic, being parameterized by the type $S$ of the stored internal value,
the type $I$ of the inputs, and the binary {\em combiner} operation
$$
\oplus: S \times I \rightarrow S.
$$
Accumulators implement two assignment
operators. Denoting with $a.\val$ the internal value of accumulator instance $a$,
\begin{itemize}
\item
  $a\ = i$ sets $a.\val$ to the provided input $i$;
\item
  $a\ \incr i$ aggregates the input $i$ into $acc.\val$ using the combiner,
  i.e. sets $a.\val$ to $a.\val \oplus i$.
\end{itemize}
Each accumulator instance has a pre-defined default for the internal value.

When an accumulator instance $a$ is referenced in a GSQL expression, it evaluates to the internally stored
value $a.val$. Therefore, the context must associate the internally stored value to the instannce of 
global accumulators (identified by name) and of vertex accumulators (identified by name and vertex).

\paragraph{Specific Accumulator Types}
We revisit the accumulator types listed in Section~\ref{sec:built-in-accums}.

{\tt SumAccum<N>} is the type of accumulators where
the internal value and input have numeric type
{\tt N}/{\tt string}
their default value is $0$/the empty string
and the combiner operation is the
arithmetic $+$/string concatenation, respectively.

{\tt MinAccum<O>} is the type of accumulators where
the internal value and input have ordered type {\tt O}
(numeric, datetime, string, vertex) 
and the combiner operation is the binary minimum function.
The default values are the (architecture-dependent)
minimum numeric value, the default date, the empty string,
and undefined, respectively. Analogously for {\tt MaxAccum<O>}.

{\tt AvgAccum<N>} stores as internal value a pair consisting of
the sum of inputs seen so far, and their count. The combiner
adds the input to the running sum and increments the count.
The default value is 0.0 (double precision).

{\tt AndAccum} stores an internal boolean value (default
{\bf true}, and takes boolean inputs, combining them using
logical conjunction.
Analogously for {\tt OrAccum}, which defaults to {\bf false}
and uses logical disjunction as combiner.

{\tt MapAccum<K,V>}
stores an internal value that is a map $m$,
where {\tt K} is the type of $m$'s keys and
{\tt V} the type of $m$'s values.
{\tt V} can itself be an accumulator type,
specifying how to aggregate values mapped by $m$ to the same key.
The default internal value is the empty map.
An input is a key-value pair $(k,v) \in K \times V$.
The combiner works as follows: if the internal map $m$
does not have an entry involving key $k$, $m$ is extended
to associate $k$ with $v$. If $k$ is already defined in $m$,
then if $V$ is not an accumulator type, $m$ is modified
to associate $k$ to $v$, overwriting $k$'s former entry.
If $V$ is an accumulator type, then $m(k)$ is an accumulator
instance. In that case $m$ is modified by
replacing $m(k)$ with the new accumulator obtained by
combining $v$ into $m(k)$ using $V$'s combiner.

\subsection{Declarations}
The semantics of a declaration is a function from contexts
to contexts.

When they are created, accumulator instances are initialized by
setting their stored internal value to a default that is defined with the accumulator type.
Alternatively, they can be initialized by explicitly setting this default value using an assignment:

$$
\mathit{type\ @n}\ = e  
$$
declares a vertex accumulator of name $n \in \vAccNames$ and type {\em type}, all of whose instances are initialized with
$\sem{\ctx}{e}$, the
result of evaluating the expression in the current context. The effect of this declaration is to create the initialized
accumulator instances and extend the current context $\ctx$ appropriately. Note that a vertex accumulator instance
is identified by the accumulator name and the vertex hosting the instance. We model this by having the context associate
the vertex accumulator name with a map from vertices to instances.
$$
\sem{}{\mathit{type\ @n}\ = e}(\ctx) = \ctx'
$$
where
$$
\ctx' = \override{\ctx}{\{@n \mapsto \bigcup_{v \in V} \{ v \mapsto \sem{\ctx}{e} \}\}}
$$
Similarly,
$$
\mathit{type\ @@n}\ = e  
$$
declares a global accumulator named $n \in \gAccNames$ of type {\em type}, whose single instance is initialized with
$\sem{\ctx}{expr}$. This instance is identified in the context simply by the accumulator name:
$$
\sem{}{\mathit{type\ @@n}\ = e}(\ctx) = \override{\ctx}{\{ @@n \mapsto \sem{\ctx}{e} \}}.
$$
Finally, global variable declarations also extend the context:
$$
\sem{}{\mathit{baseType\ var}\ = e}(\ctx) = \override{\ctx}{\{ var \mapsto \sem{\ctx}{e} \}}.
$$

\subsection{\darpe\ Semantics}
\darpe s specify a set of paths in the graph, formalized as follows.

\paragraph{Paths}
A {\em path} $p$ in graph $G$ is a sequence
$$
p = v_0,\ e_1,\ v_1,\ e_2,\ v_2,\ \ldots,\ v_{n-1},\ e_n,\ v_n
$$
where $v_0 \in V$ and for each $1 \leq i \leq n$,
\begin{itemize}
\item
  $v_i \in V$, and
\item
  $e_i \in E$, and
\item
  $v_{i-1}$ and $v_i$ are the endpoints of edge $e_i$ regardless of $e_i$'s
  orientation: $(v_{i-1},v_i) \in \epts(e_i)$ or $(v_i,v_{i-1}) \in \epts(e_i)$.
\end{itemize}

\paragraph{Path Length}
We call $n$ the {\em length} of $p$ and denote it with $\length{p}$.
Note that when $p=v_0$ we have $\length{p} = 0$.

\paragraph{Path Source and Target}
We call $v_0$ the {\em source},
and $v_n$ the {\em target} of path $p$, denoted $\src(p)$ and $\tgt(p)$,
respectively. When $n=0$, $\src(p) = \tgt(p) = v_0$.

\paragraph{Path Hop}
For each $1 \leq i \leq n$, we call the triple $(v_{i-1}, e_i, v_i)$ the {\em hop $i$} of path $p$.

\paragraph{Path Label}
We define the {\em label of hop $i$ in $p$},
denoted $\lambda_i(p)$ as follows:
$$
\lambda_i(p) =
\left\{
\begin{array}{l@{,\ \ \mbox{if\ }}l}
  \tyE(e_i)       & \mbox{$e_i$ is undirected}, \\
  \tyE(e_i)>  & \mbox{$e_i$ is directed from $v_{i-1}$ to $v_i$}, \\
  <\tyE(e_i)  & \mbox{$e_i$ is directed from $v_i$ to $v_{i-1}$}
\end{array} 
\right.
$$
where
$\tyE(e_i)$ denotes the type name of edge $e_i$,
and $\tyE(e_i)>$ denotes a new symbol obtained by concatenating
$\tyE(e_i)$ with $>$, and analogously for $<\tyE(e_i)$.

We call {\em label} of p, denoted \lab(p), the word obtained by
concatenating the hop labels of $p$:
$$
\lab (p) =
\left\{
\begin{array}{l@{,\ \ \mbox{if\ }}l}
  \epsilon & \length{p} = 0,\\
  \lab_1(p)\lab_2(p)\ldots \lab_{\length{p}}(p) & \length{p} > 0
\end{array}
\right.
$$
where, as usual~\cite{hopcroft-ullman}, $\epsilon$ denotes the empty word.

\paragraph{\darpe\ Satisfaction}
We denote with
$$
\daAlpha = \bigcup_{t \in \eTyNames} \{ t, t>, <t \}
$$
the set of symbols obtained by treating each edge type $t$ as a symbol,
as well as creating new symbols by concatenating $t$ and $>$, as well as $<$ and $t$.

We say that path $p$ {\em satisfies} \darpe\ D, denoted
$p \models D$, if
$\lab(p)$ is a word in the language accepted by D, $\calL(D)$,
when viewed as a regular expression over the alphabet $\daAlpha$~\cite{hopcroft-ullman}:
$$
p \models D\ \Leftrightarrow\ \lab(p) \in \calL(D).
$$

\paragraph{\darpe\ Match}
We say that \darpe\ D\ {\em matches} path $p$ (and that $p$ {\em is a match for}
D) whenever $p$ is a shortest path that satisfies D:
$$
p \mbox{\ matches\ } D \Leftrightarrow\  p \models D\ \ \wedge\ \
\length{p} = \min_{q \models D} \length{q}.
$$
We denote with \sem{G}{D}
the set of matches of a \darpe\ $D$ in graph $G$.

\subsection{Pattern Semantics}
Patterns consist of \darpe s and variables.
The former specify a set of paths in the graph, the latter are bound to
vertices/edges occuring on these paths.

A pattern $P$ specifies a function from a graph $G$ and a context $\ctx$ 
to
\begin{itemize}
  \item
    a set $\sem{G,\ctx}{P}$ of paths in the graph, each called a {\em match} of  $P$, and
  \item
    a family $\{\binding{P}{p}\}_{p \in \sem{G,\ctx}{P}}$ of bindings for $P$'s variables,
one for each match $p$. Here, $\binding{P}{p}$ denotes the binding {\em induced} by match $p$. 
\end{itemize}

\paragraph{Temporary Tables and Vertex Sets}
To formalize pattern semantics, we note that
some GSQL query statments may construct temporary tables that can be referred to by subsequent
statements.
Therefore, among others, the context maps the names to the extents (contents) of temporary tables.
Since we can model a set of vertices as a single-column, duplicate-free table containining
vertex ids, we refer to such tables as {\em vertex sets}.

\paragraph{V-Test Match}
Consider graph $G$ and a context \ctx.
Given a v-test $VT$, a {\em match}  for $VT$ 
is a vertex $v \in V$ (a path of length 0) such that
$v$ belongs to $VT$ (if $VT$ is a vertex set name defined in \ctx),
or $v$ is a vertex of type $VT$ (otherwise).
We denote the set of all matches of $VT$ against $G$ in context \ctx\
with $\sem{G,\ctx}{VT}$.
$$
v \in \sem{G,\ctx}{VT} \Leftrightarrow \left\{
\begin{array}{l@{\ \ }l}
  v \in \ctx(VT), & \mathit{if}\ VT\ \mathit{is\ a\ vertex\ set}\\
                 & \mathit{defined\ in\ } \ctx,\\
  \tyV(v) = VT,   & \mathit{if}\ VT \in \dom(\tyV)\ \mathit{i.e.}\\
                 &  \mathit{VT\ is\ a\ defined\ vertex\ type}
\end{array}
\right.
$$

\paragraph{Variable Bindings}
Given graph $G$ and tuple of variables ${\bf x}$,
a {\em binding} for ${\bf x}$ in $G$ is a function $\beta$ from the variables in ${\bf x}$
to vertices or edges in $G$, $\beta: {\bf x} \rightarrow V \cup E$.
Notice that a variable binding (binding for short) is a particular kind of context, hence
all context-specific definitions and operators apply. In particular, the notion of
consistent bindings coincides with that of consistent contexts.

\paragraph{Binding Tables}
We refer to a bag of variable bindings as a {\em binding table}.

\paragraph{No-hop Path Pattern Match}
Given a graph $G$ and a context \ctx,
we say that path $p$ is a match
for no-hop pattern $P = VT:x$
(and we say that $P$ {\em matches} $p$)
if $p \in \sem{G,\ctx}{VT}$.
Note that $p$ is a path of length 0, i.e. a vertex $v$.
The match $p$ {\em induces a binding} of vertex variable $x$ to $v$,
$\binding{P}{p} = \{ x \mapsto v \}$.
~\footnote{
Note that both $VT$ and $x$ are optional.
If $VT$ is missing, then it is trivially satisfied by all vertices.
If $x$ is missing, then the induced binding is the emtpy map $\binding{P}{p} = \{ \}$.
These conventions apply for the remainder of the presentation and are not repeated
explicitly.}

\paragraph{One-hop Path Pattern Match}
Recall that in a one-hop pattern
$$
P = S:s -(D:e)- T:t,
$$
$D$ is a disjunction of direction-adorned edge types, $D \subseteq \daAlpha$,
$S,T$ are v-tests, $s,t$ are vertex variables and $e$ is an edge variable.
We say that (single-edge) path $p = v_0,\ e_1,\  v_1$ {\em is a match for} $P$
(and that $P$ {\em matches} $p$) if
  $v_0 \in \sem{G,\ctx}{S}$, and
  $p \in \sem{G}{D}$, and
  $v_1 \in \sem{G,\ctx}{T}$.
The {\em binding induced} by this match, denoted $\binding{P}{p}$, is
$\binding{P}{p} = \{ s \mapsto v_0, t \mapsto v_1, e \mapsto e_1 \}$.

\paragraph{Multi-hop Single-\darpe\ Path Pattern Match}
Given a \darpe\ $D$, path $p$ {\em is a match for}
multi-hop path pattern $P = S:s -(D)- T:t$ (and $P$ {\em matches} $p$) if
  $p \in \sem{G}{D}$, and
  $\src(p) \in \sem{G,\ctx}{S}$ and
  $\tgt(p) \in \sem{G,\ctx}{T}$.
The match induces a {\em binding}
$\binding{P}{p} = \{ s \mapsto \src(p), t \mapsto \tgt(p) \}$.

\paragraph{Multi-\darpe\ Path Pattern Match}
Given \darpe s $\{D_i\}_{1\leq i \leq n}$, {\em a match for}
path pattern
$$
P = S_0:s_0 -(D_1)- S_1:s_1 - \ldots -(D_n)- S_n:s_n
$$
is a tuple of segments $(p_1, p_2, \ldots, p_n)$
of a path $p = p_1p_2\ldots p_n$ such that
$p \in \sem{G}{D_1.D_2.\ldots.D_n}$, and for each $1 \leq i \leq n$:
$p_i \in \sem{G,\ctx}{S_{i-1}:s_{i-1} -(D_i)- S_i:s_i}$.
Notice that
$p$ is a shortest path satisfying
the \darpe\ obtained by concatenating the \darpe s of the pattern.
Also notice that there may be multiple matches that correspond to distinct segmenations
of the same path $p$. Each match induces a binding
$$
\binding{P}{p_1.\ldots.p_n} = \bigcup_{i=1}^n \binding{S_{i-1}:s_{i-1} -(D_i)- S_i:s_i}{p_i}.
$$
Notice that the individual bindings induced by the segments are pairwise consistent, since the
target vertex of a path segment coincides with the source of the subsequent segment.

\paragraph{Consistent Matches}
Given two patterns $P_1, P_2$ with matches $p_1, p_2$ respectively,
we say that $p_1, p_2$ are {\em consistent} if the bindings they induce are consistent.
Recall that bindings are contexts so they inherit the notion of consistency.

\paragraph{Conjunctive Pattern Match}
Given a conjunctive pattern ${\bf P} = P_1, P_2, \ldots, P_n$,
a {\em match} of $\bf P$ is a tuple of paths ${\bf p} = p_1, p_2, \ldots, p_n$
such that $p_i$ is a match of $P_i$ for each $1 \leq i \leq n$ and
all matches are pairwise consistent. ${\bf p}$ induces a binding on all
variables occuring in ${\bf P}$,
$$
\binding{{\bf P}}{{\bf p}} = \bigcup_{i=1}^n \binding{P_i}{p_i}.
$$

\subsection{Atom Semantics}
Recall that the FROM clause comprises a sequence of relational or graph atoms.
Atoms specify a pattern and a collection to match it against. Each match induces a variable binding.
Note that multiple distinct matches of a pattern can induce the same variable binding if they
only differ in path elements that are not bound to the variables.
To preserve the multiplicities of matches, we define the semantics of atoms
a function from contexts to binding tables, i.e. bags of bindings
for the variables introduced in the pattern.

Given relational atom $T\ \AS\ x$ where $T$ is a table name and $x$ a tuple variable,
the matches of $x$ are the tuples $t$ from $T$'s extent, $t \in \ctx(T)$, and they
each induce a binding $\binding{x}{t} = \{ x \mapsto t \}$.
The semantics of $T\ \AS\ x$ is the function
$$
\sem{\ctx}{T\ \AS\ x} = \biguplus_{t \in \ctx(T)} \bag{\binding{x}{t}}.
$$
Here, $\bag{x}$ denotes the singleton bag containing element $x$
with multiplicity 1, and $\biguplus$ denotes bag union, which is multiplicity-preserving
(like SQL's UNION ALL operator).

Given graph atom $G\ \AS\ P$ where $P$ is a conjunctive pattern, its
semantics is the function
$$
\sem{\ctx}{G\ \AS\ P} = \biguplus_{p \in \sem{\ctx(G),\ctx}{P}} \bag{\binding{P}{p}}.
$$
When the graph is not explicitly specified in the atom, we use the default graph $DG$,
which is guaranteed to be set in the context:
$$
\sem{\ctx}{P} = \biguplus_{p \in \sem{\ctx(DG),\ctx}{P}} \bag{\binding{P}{p}}.
$$

\subsection{FROM Clause Semantics}
The \from\ clause specifies a function from contexts to (context, binding table) pairs
that preserves the context:
$$
\sem{}{\from\ a_1,\ldots,a_n}(\ctx)
$$
$$
= (\ctx, 
\biguplus_{
  \begin{array}{c}
    \beta_1 \in \sem{\ctx}{a_1},\ldots,\beta_n \in \sem{\ctx}{a_n},\\
    \ctx,\beta_1,\ldots,\beta_n\ \mathit{pairwise\ consistent}
  \end{array}
} \bag{\bigcup_{i=1}^n \beta_i}).
$$
    
The requirement that all $\beta_i$s be pairwise consistent
addresses the case when atoms share variables (thus implementing joins).
The consistency of each $\beta_i$ with \ctx\ covers the case when
the pattern $P$ mentions variables defined prior to the current query block.
These are guaranteed to be defined in \ctx\ if the query passes the semantic checks.

\subsection{Term Semantics}
Note that GSQL terms conform to SQL syntax, extended to
also allow accumulator- and type-specific terms of form
\begin{itemize}
\item
  $@@A$, referring to the value of global accumulator $A$,
\item
  $@@A'$, referring to the value of global accumulator $A$ prior to the execution of the current
  query block,
\item
  $x.@A$, specifying the value of the vertex accumulator $A$ at the vertex denoted by variable $x$,
\item
  $x.@A'$, specifying the value of the vertex accumulator $A$ at the vertex denoted by variable $x$
  prior to the execution of the current query block,
\item
  $\ty{x}$, referring to the type of the vertex/edge denoted by variable $x$.
\end{itemize}

The variable $x$ may be a query parameter,
or a global variable set by a previous statement, or a variable
introduced by the \from\ clause pattern. Either way, its value
must be given by the context \ctx\ assuming that $x$ is defined.

A constant $c$ evaluates to itself:
$$
\sem{\ctx}{c} = c.
$$
A variable evaluates to the value it is bound to in the context:
$$
\sem{\ctx}{var} = \ctx(var).
$$
An attribute projection term $x.A$ evaluates to the value of the $A$ attribute of $\ctx(x)$:
$$
\sem{\ctx}{x.A} = \data(\ctx(x),A).
$$

The evaluation of the accumulator-specific terms requires the context \ctx\ to also
\begin{itemize}
  \item
    map global accumulator names to their values, and
  \item
    map each vertex accumulator name (and the primed version referring to the prior value)
    to a map from vertices to accumulator values,
    to model the fact that each vertex has its own accumulator instance.
\end{itemize}
Given context \ctx,
terms of form $@@A$ evaluate to the value of the global accumulator as recorded in the context:
$$
\sem{\ctx}{@@A} = \ctx(@@A).
$$
Terms of form $@@A'$ evaluate to the prior value of the global accumulator instance,
as recorded in the context:
$$
\sem{\ctx}{@@A'} = \ctx(@@A').
$$
Terms of form $x.@A$ evaluate to the value of the vertex accumulator instance located at the
vertex denoted by variable $x$:
$$
\sem{\ctx}{x.@A} = \ctx(@A)(\ctx(x)).
$$
Terms of form $x.@A'$ evaluate to the prior value of the vertex accumulator instance located at the
vertex denoted by variable $x$:
$$
\sem{\ctx}{x.@A'} = \ctx(@A')(\ctx(x)).
$$
Analogously for global accumulators:
$$
\sem{\ctx}{@@A'} = \ctx(@@A').
$$
All above definitions apply when $A$ is not of type {\tt AvgAccum}, which is treated
as an exception: in order to support order-invariant implementation, the value
associated in the context is a (sum,count) pair and terms evaluate to the
division of the sum component by the count component. If $A$ is of type {\tt AvgAccum},
$$
\sem{\ctx}{@@A} = s/c,\ \mathit{where\ } (s,c) = \ctx(@@A)
$$
and
$$
\sem{\ctx}{x.@A} = s/c, \mathit{where\ } (s,c) = \ctx(@A)(\ctx(x)).
$$

Finally, terms of form $\ty{x}$ evaluate as follows:
$$
\sem{\ctx}{\ty{x}} = \left\{\begin{array}{ll}
  \tyV(\ctx(x)) & \mathit{if\ x\ is\ vertex\ variable}\\
  \tyE(\ctx(x)) & \mathit{if\ x\ is\ edge\ variable}
\end{array}
\right.
$$

\subsection{Expression Semantics}\label{sec:expr-semantics}
The semantics of GSQL expressions is compatible with
that of SQL expressions.
Expression evaluation is defined in the standard SQL way by induction on their
syntactic structure, with the evaluation of terms as the base case.
We denote with
$$
\sem{\ctx}{E}
$$
the result of evaluating
expression $E$ in context \ctx.
Note that this definition also covers conditions (e.g. such as used in the \where\ clause)
as they are particular cases (boolean expressions).

\subsection{\where\ Clause Semantics}
The semantics of a where clause 
$$
\where\ \mathit{Cond}
$$
is a function
$
\sem{}{\where\ \mathit{Cond}}
$
on (context, binding table)-pairs that preserves the context.
It filters the input bag $B$ keeping only the bindings that satisfy condition
$\mathit{Cond}$ (preserving their multiplicity):
$$
\sem{}{\where\ \mathit{Cond}}(\ctx,B) = 
(\ctx,
\biguplus_{\begin{array}{c}\beta \in B,\\ \sem{\override{\ctx}{\beta}}{\mathit{Cond}} = \mathit{true}\end{array}} \bag{\beta}
).
$$
Notice that the condition is evaluated for each binding $\beta$
in the new context $\override{\ctx}{\beta}$
obtained by extending \ctx\ with $\beta$.

\subsection{ACCUM Clause Semantics}
The effect of the \accum\ clause is to modify accumulator
values.  It is executed exactly once for every variable binding $\beta$
yielded by the \where\ clause (respecting multiplicities).
We call each individual execution of the \accum\ clause an {\em acc-execution}.

Since multiple acc-executions may refer to the same accumulator
instance, they do not write  the accumulator value directly,
to avoid setting up race conditions in which acc-executions overwrite each other's writes
non-deterministically. Instead, each acc-execution yields a bag of input
values for the accumulators mentioned in the \accum\ clause.
The cumulative effect of all acc-executions
is to aggregate all generated inputs into the 
appropriate accumulators, using the accumulator's $\oplus$ combiner.

\paragraph{Snapshot Semantics}
Note that all acc-executions start from the same snapshot of accumulator values and
the effect of the accumulator inputs produced by each acc-execution
are not visible to the acc-executions.
These inputs are aggregated into accumulators
only after all acc-executions have completed.
We therefore say that the \accum\ clause executes under {\em snapshot semantics},
and we conceptually structure this execution into two phases:
in the {\em Acc-Input Generation Phase}, all acc-executions compute accumulator inputs
(acc-inputs for short), starting from the same accumulator value snapshot.
After all acc-executions complete, the {\em Acc-Aggregation Phase} aggregates the generated inputs into the accumulators they are
destined for.
The result of the acc-aggregation phase is a new snapshot of the accumulator values.

\subsubsection{Acc-Input Generation Phase}\ \\
To formalize the semantics of this phase, we denote with
$\bag{\calD}$ the class of bags with elements from \calD.

We introduce
two new maps that associate accumulators with the bags of inputs
produced for them during the acc-execution phase:
\begin{itemize}
\item
  $\deltaG : \gAccNames \rightarrow \bag{\calD}$,
  a {\em global acc-input map} that associates global accumulators instances (identified by name)
  with a bag of input values. 
\item
  $\deltaV : (V \times \vAccNames) \rightarrow \bag{\calD}$,
  a {\em vertex acc-input map} that associates vertex accumulator instances (identified by (vertex,name) pairs)
  with a bag of input values. 
\end{itemize}
For presentation convenience, we regard both maps as total functions (defined on all
possible accumulator names) with finite support (only a finite number of accumulator names
are associated with a non-empty acc-input bag).

\paragraph{Acc-statements}
An \accum\ clause consists of a sequence of
statements
$$
\accum\ s_1,s_2,\ldots,s_n
$$
which we refer to as \astmt s to distinguish
them from the statements appearing outside the \accum\ clause.

\paragraph{Acc-snapshots}
An {\em acc-snapshot} is a tuple
$$
(\ctx, \deltaG, \deltaV)
$$
consisting of
  a context \ctx,
  a global acc-input map \deltaG, and
  a vertex acc-input map \deltaV.

\paragraph{Acc-statement semantics}
The semantics of an \astmt\ $s$ is a function $\sem{}{s}$
from acc-snapshots to acc-snapshots.
$$
(\ctx, \deltaG, \deltaV)
\stackrel{\sem{}{s}}{\longrightarrow}
(\ctx', \deltaG', \deltaV').
$$

\paragraph{Local Variable Assignment}
Acc-statements may introduce new {\em local variables}, whose scope
is the remainder of the ACCUM clause, or they may assign values to such local variables. 
Such acc-statements have form
$$
\mathit{type?\ lvar\ = expr}
$$
If the variable was already defined, then the type specification is missing
and the acc-statement just updates the local variable.

The 
semantics of local variable assignments and declarations is a function that extends (possibly overriding)
the context with a binding of local variable $\mathit{lvar}$
to the result of evaluating $\mathit{expr}$:
$$
\sem{}{\mathit{type\ lvar\ = expr}}_\eta (\ctx, \deltaG, \deltaV)=
(\ctx', \deltaG, \deltaV)$$
where
$$
\ctx' =
\override{\ctx}{\{\mathit{lvar}\ \mapsto \sem{\ctx}{\mathit{expr}}\}}.
$$

\paragraph{Input to Global Accums}
Acc-statements that input into a global accumulator have form
$$
@@A \incr \mathit{expr}
$$
and their semantics is a function that adds the evaluation of
expression $\mathit{expr}$ to the bag of inputs for accumulator $A$, $\deltaG(A)$:
$$
\sem{}{\mathit{@@A\ \incr expr}}_\eta (\ctx, \deltaG, \deltaV)=
(\ctx, \deltaG', \deltaV)$$
where
\begin{eqnarray*}
\deltaG' & = &
\override{\deltaG}{\{A\ \mapsto \mathit{val} \}}\\
\mathit{val} & = &
\deltaG(A) \uplus \bag{\sem{\ctx}{\mathit{expr}}}.
\end{eqnarray*}

\paragraph{Input to Vertex Accum}
Acc-statements that input into a vertex accumulator have form
$$
x.@A \incr \mathit{expr}
$$
and their semantics is a function that adds the evaluation of
expression $\mathit{expr}$ to the bag of inputs for the instance of accumulator $A$
located at the vertex denoted by $x$ (the vertex is $\ctx(x)$ and the bag of inputs
is $\deltaV(\ctx(x),A)$):
$$
\sem{}{\mathit{x.@A \incr expr}}_\eta (\ctx, \deltaG, \deltaV)=
(\ctx, \deltaG, \deltaV')
$$
where
\begin{eqnarray*}
\deltaV' & = &
\override{\deltaV}{\{
(\ctx(x), A)\ \mapsto \mathit{val} \}}\\
\mathit{val} & = &
\deltaV(\ctx(x), A) \uplus \bag{\sem{\ctx}{\mathit{expr}}}.
\end{eqnarray*}

\paragraph{Control Flow}
Control-flow statements, such as

$\emph{{\tt if}\ cond\ {\tt then}\ \astmt s\ {\tt else}\ \astmt s\ {\tt end}}$\\
and

$\emph{{\tt foreach}\ var\ {\tt in}\ expr\ {\tt do}\ \astmt s\ {\tt end}}$\\
etc., evaluate in the standard fashion of structured programming languages.

\paragraph{Sequences of \astmt s}
The semantics of a sequence of \astmt s
is the standard composition of the functions
specified by the semantics of the individual \astmt s:

$$
\sem{}{s_1,s_2,\ldots,s_n}_\eta = \sem{}{s_1}_\eta \circ \sem{}{s_2}_\eta \circ \ldots \circ \sem{}{s_n}_\eta. 
$$
Note that since the \astmt s do not directly modify the value of accumulators,
they each evaluate using the same snapshot of accumulator values.

\paragraph{Semantics of Acc-input Generation Phase}
We are now ready to formalize the semantics of the acc-input generation phase as
a function
$$
\sem{}{\accum\ s_1,s_2,\ldots,s_n}_\eta
$$
that takes as input
a context \ctx\ and a binding table $B$ and returns a pair of (global and
vertex) acc-input maps that associate to each accumulator their cumulated bag of acc-inputs.
See Figure~\ref{fig:acc-input-generation}, where
$\deltaG^{\bag{}}$ denotes the map assigning to each global accumulator name
the empty bag, and
$\deltaV^{\bag{}}$ denotes the map assigning to each
(vertex, vertex accumulator name) pair the empty bag.

\begin{figure*}
\begin{minipage}{\textwidth}
\begin{eqnarray*}
  \sem{}{\accum\ ss}_\eta(\ctx, B) & = & (\deltaG', \deltaV')\\\\
  \deltaG' & = & \bigcup_{A \mathit{\ global\ acc\ name}}
  \{A \mapsto \biguplus_{\begin{array}{c}\beta \in B,\\ (\_, \deltaG, \_) = \sem{}{ss}_\eta(\override{\ctx}{\beta}, \deltaG^{\bag{}}, \deltaV^{\bag{}})\end{array}} \deltaG(A) \}\\\\\\
  \deltaV' & = & \bigcup_{v \in V,\ A \mathit{\ vertex\ acc\ name}}
  \{(v, A) \mapsto \biguplus_{\begin{array}{c}\beta \in B,\\ (\_, \_, \deltaV) = \sem{}{ss}_\eta(\override{\ctx}{\beta}, \deltaG^{\bag{}}, \deltaV^{\bag{}})\end{array}} \deltaV(v, A) \}. 
\end{eqnarray*}
\end{minipage}
\caption{The Semantics of the Acc-Input Generation Phase Is a Function $\sem{}{}_\eta$}
\label{fig:acc-input-generation}
\end{figure*}

\subsubsection{Acc-Aggregation Phase}\ \\
This phase aggregates the generated acc-inputs into the accumulator they are meant for.

We formalize this effect using the \reduce\ function.
It is parameterized by a binary operator $\oplus$ and it takes as inputs
an intermediate value $v$ and a bag $B$ of inputs and returns the result of
repeatedly $\oplus$-combining $v$ with the inputs in $B$, in non-deterministic
order.

$$
\reduce{\oplus} (v, B) =
\left\{ \begin{array}{l@{,\ \ \mathit{if }\ }l}
  v & B = \bag{}\\
  \reduce{\oplus} (v \oplus i, B - \bag{i}) & i \in B
\end{array} \right.  
$$
Note that operator $-$ denotes the standard bag difference, whose effect here is to
decrement by 1 the multiplicity of $i$ in the resulting bag.
Also note that the choice of acc-input $i$ in bag $B$ is not further specified,
being non-deterministic.

Figure~\ref{fig:acc-aggregation} shows how the \reduce\ function is used for
each accumulator to yield a new accumulator value that incorporates
the accumulator inputs. The resulting semantics of the aggregation phase is a function
$\reduce{\mathit{all}}$
that takes as input the context, the global and vertex acc-input maps,
and outputs a new context reflecting the new accumulator values,
also updating the prior accumulator values.
\begin{figure*}
\begin{minipage}{\textwidth}
\begin{eqnarray*}
  \reduce{\mathit{all}} (\ctx,\deltaG,\deltaV) & = &
  \ctx'\\
  \ctx' & = &
  \override{(\override{\ctx}{\mapGAcc})}{\mapVAcc}\\
  \mapGAcc & = &
  \bigcup_{A\ \in \gAccNames}\{A' \mapsto \ctx(A),\ \ A \mapsto \reduce{A.\oplus}(\ctx(A), \deltaG(A))\}\\
  \mapVAcc & = &
  \bigcup_{A\ \in \vAccNames} \{
  A' \mapsto
  \bigcup_{v \in V}\{v \mapsto \ctx(A)(v)\},\ \ 
  A \mapsto 
  \bigcup_{v \in V}\{v \mapsto \reduce{A.\oplus}(\ctx(A)(v), \deltaV(v,A))\}
  \}
\end{eqnarray*}
\end{minipage}
\caption{The Semantics of the Acc-Aggregation Phase Is a Function $\reduce{\mathit{all}}$}
\label{fig:acc-aggregation}
\end{figure*}

\paragraph{Order-invariance}
The result of the acc-aggregation phase is well-defined (input-order-invariant) for an accumulator
instance $a$ whenever the binary aggregation operation $a.\oplus$ is commutative and associative.
This is the case for built-in GSQL accumulator types
SetAccum, BagAccum, HeapAccum, OrAccum, AndAccum, MaxAccum, MinAccum, SumAccum, 
and even AvgAccum (implemented in an order-invariant way by having the stored internal value be
the pair of sum and count of inputs). Order-invariance also holds for complex accumulator types
MapAccum and GroupByAccum if they are based on order-invariant accumulators.
The exceptions are the List-, Array- and StringAccm accumulator types.

\subsubsection{Putting It All Together}\ \\
The semantics of the \accum\ clause is a function on
(context, binding table)-pairs that preserves the binding table:
$$
\sem{}{\accum\ s_1,s_2,\ldots,s_n}(\ctx, B)
= (\ctx', B)
$$
where
$$
\ctx' = \reduce{\mathit{all}} (\ctx, \deltaG, \deltaV)
$$
with $\deltaG,\deltaV$ provided
by the acc-input generation phase:
$$
(\deltaG,\deltaV) =
\sem{}{\accum\ s_1,s_2,\ldots,s_n}_\eta(\ctx, B).
$$

\subsection{\paccum\ Clause Semantics}
The purpose of the \paccum\ clause is to specify
computation that takes place after accumulator values
have been set by the \accum\ clause (at this point, the effect
of the Acc-Aggregation Phase is visible).

The computation is specified by a sequence of acc-statements,
whose syntax and semantics coincide with that of the sequence of
acc-statements in an \accum\ clause:

$$
\sem{}{\paccum\ ss} = \sem{}{\accum\ ss}.
$$

The \paccum\ clause does not add expressive power to GSQL,
it is syntactic sugar introduced for convenience and conciseness.

\subsection{SELECT Clause Semantics}
The semantics of the GSQL \select\ clause is modeled after standard SQL.
It is a function on (context, binding table)-pairs that preserves the context
and returns a table whose tuple components are the results of evaluating the \select\ clause 
expressions as described in Section~\ref{sec:expr-semantics}.

\paragraph{Vertex Set Convention}
We model a vertex set as a single-column,
duplicate-free table whose values are vertex ids.
Driven by our experience with GSQL customer deployments,
the default interpretation of clauses of form
$
\select\ x
$
where $x$ is a vertex variable is
$
\select\ \distinct\ x.
$
To force the output of a vertex bag, the
developer may use the keyword ALL:
$
\select\ \mbox{\ ALL}\ x.
$

Standard SQL bag semantics applies in all other cases.

\subsection{GROUP BY Clause Semantics}
GSQL's GROUP BY clause follows standard SQL semantics.
It is a function on (context, table)-pairs that preserves the context.
Each group is represented as a nested-table component in a tuple
whose scalar components hold the
group key.

\subsection{HAVING Clause Semantics}
GSQL's HAVING clause follows standard SQL semantics.
It is a function from (context,table) pairs to tables.

\subsection{ORDER BY Clause Semantics}
GSQL's ORDER BY clause follows standard SQL semantics.
It is a function from (context,table) pairs to ordered tables (lists of tuples).

\subsection{LIMIT Clause Semantics}
The LIMIT clause inherits standard SQL's semantics.
It is a function from (context,table) pairs to tables.
The input table may be ordered, in which case the output table is obtained
by limiting the prefix of the tuple list. If the input table is unordered (a bag),
the selection is non-deterministic.

\subsection{Query Block Statement Semantics}
Let $qb$ be the query block
\begin{lstlisting}
   SELECT (*@{\tt\textbf{DISTINCT}}@*)? s INTO t
   FROM       f
   WHERE      w
   ACCUM      a
   POST-ACCUM p
   GROUP BY   g
   (*@{\tt\textbf{HAVING}}@*)     h
   ORDER BY   o
   LIMIT      l.
\end{lstlisting}
Its semantics is a function on contexts, given by the composition of the
semantics of the individual clauses:
\begin{eqnarray*}
  \sem{}{qb}(\ctx) & = & \override{\ctx'}{\{t \mapsto T \}}\\
  \mathit{where}\\
  (\ctx', T)  & = & \sem{}{\from\ f}\\
              & \circ & \sem{}{\where\ w}\\
              & \circ & \sem{}{\accum\ a}\\
              & \circ & \sem{}{\groupby\ g}\\
              & \circ & \sem{}{\select\ \distinct?\ s}\\
              & \circ & \sem{}{\paccum\ p}\\
              & \circ & \sem{}{\having\ h}\\
              & \circ & \sem{}{\orderby\ o}\\
              & \circ & \sem{}{\limit\ l} (\ctx)
\end{eqnarray*}

\paragraph{Multi-Output \select\ Clause}
Multi-Output queries have the form
\begin{lstlisting}
   SELECT     (*@$s_1$@*) INTO (*@$t_1$@*); ... ; (*@$s_n$@*) INTO (*@$t_n$@*)
   FROM       f
   WHERE      w
   ACCUM      a
   GROUP BY   (*@$g_1$@*); ... ; (*@$g_n$@*)
   (*@{\tt\textbf{HAVING}}@*)     (*@$h_1$@*); ... ; (*@$h_n$@*)
   LIMIT      (*@$l_1$@*); ... ; (*@$l_n$@*).
\end{lstlisting}
They output several tables based on the same \from, \where, and \accum\ clause.
Notice the missing \paccum\ clause.
The semantics is also a function on contexts, where the output context reflects
the multiple output tables:
\begin{eqnarray*}
  \sem{}{qb}(\ctx) & = & \override{\ctx_A}{\{t_1 \mapsto B_1, \ldots, t_n \mapsto B_n \}} \\
  \mathit{where}\\
  (\ctx_A,B_A)  & = & \sem{}{\from\ f}\\
                & \circ & \sem{}{\where\ w}\\
                & \circ & \sem{}{\accum\ a}(\ctx)\\
  \\
                &   & \mathit{for\ each}\ 1 \leq i \leq n\\
  (\ctx_A, B_i)  & = & \sem{}{\groupby\ g_i}\\
                & \circ & \sem{}{\select\ s_i}\\
                & \circ & \sem{}{\having\ h_i}\\
                & \circ & \sem{}{\limit\ l_i}(\ctx_A, B_A)\\
\end{eqnarray*}

\subsection{Assignment Statement Semantics}
The semantics of assignment statements is a function from contexts to contexts.\\

For statements that introduce new global variables, or update previously introduced global variables,
their semantics is
$$
\sem{}{type?\ gvar\ = expr} (\ctx) = \override{\ctx}{\{gvar \mapsto \sem{\ctx}{expr}\}}.
$$

For statements that assign to a global accumulator, the semantics is
$$
\sem{}{@@gAcc\ = expr} (\ctx) = \override{\ctx}{\{@@gAcc \mapsto \sem{\ctx}{expr}\}}.
$$

\paragraph{Vertex Set Manipulation Statements} In the following, let $s, s_1, s_2$
denote vertex set names.
$$
\sem{}{s = expr} (\ctx) = \override{\ctx}{\{s \mapsto \sem{\ctx}{expr}\}}.
$$
$$
\sem{}{s = s_1\ {\tt union}\ s_2} (\ctx) = \override{\ctx}{\{s \mapsto \ctx(s_1)\cup\ctx(s_2)\}}.
$$
$$
\sem{}{s = s_1\ {\tt intersect}\ s_2} (\ctx) = \override{\ctx}{\{s \mapsto \ctx(s_1)\cap\ctx(s_2)\}}.
$$
$$
\sem{}{s = s_1\ {\tt minus}\ s_2} (\ctx) = \override{\ctx}{\{s \mapsto \ctx(s_1)-\ctx(s_2)\}}.
$$

\subsection{Sequence of Statements Semantics}
The semantics of statements is a function from contexts to contexts.
It is the composition of the individual statement semantics:
$$
\sem{}{s_1 \ldots s_n} = \sem{}{s_1} \circ \ldots \circ \sem{}{s_n}.
$$

\subsection{Control Flow Statement Semantics}
The semantics of control flow statements is a function from contexts to contexts,
conforming to the standard semantics of structured programming languages.
We illustrate on two examples.

\paragraph{Branching statements}
$$
\sem{}{{\tt if}\ cond\ {\tt then}\ stmts_1\ {\tt else}\ stmts_2\ {\tt end}}(\ctx) 
$$
$$
\ \ \
= \left\{
\begin{array}{l@{,\ \ \mathit{if}\ }l}
    \sem{}{stmts_1}(\ctx) & \sem{\ctx}{cond} = \mathit{true}\\
    \sem{}{stmts_2}(\ctx) & \sem{\ctx}{cond} = \mathit{false}    
\end{array}
\right.
$$

\paragraph{Loops}\ \\
\noindent
$
\sem{}{{\tt while}\ cond\ {\tt do}\ stmts\ {\tt end}}(\ctx)
$

\noindent
$$
\ \ \
= \left\{
\begin{array}{ll}
    \ctx, & \mathit{if}\ \sem{\ctx}{cond} = \mathit{false}\\    
    \sem{}{stmts} \circ\\
    \sem{}{{\tt while}\ cond\ {\tt do}\ stmts\ {\tt end}}(\ctx), & \mathit{if}\  \sem{\ctx}{cond} = \mathit{true}\\
\end{array}
\right.
$$

\subsection{Query Semantics}
A query $Q$ that takes $n$ parameters is a function
$$
Q: \{ \calD \mapsto \calD \} \times \calD^n \rightarrow \calD
$$
where $\{ \calD \mapsto \calD \}$ denotes the class of maps (meant to model contexts that map graph names to their contents),
and $\calD^n$ denotes the class of $n$-tuples (meant to provide arguments for the parameters).

Consider query $Q$ below, where the $p_i$'s denote parameters, the $d_i$s denote declarations,
the $s_i$s denote statements and the $a_i$s denote arguments instantiating the parameters.
Its semantics is the function
\begin{tabbing}
  \\
  $[\![$ \= ${\tt create}$\ \= ${\tt query\ }Q\ (ty_1\ p_1, \ldots, ty_n\ p_n)\ \{$\+\+\\
      $d_1; \ldots; d_m;$\\
      $s_1; \ldots; s_k;$\\
      ${\tt return}\ e;$\-\\
      $\}$\-\\
      $]\!]\ (\ctx,a_1,\ldots,a_n) = \sem{\ctx'}{e}$\\
\end{tabbing}
where
\begin{eqnarray*}
  \ctx' & = & \sem{}{d_1} \circ \ldots \circ \sem{}{d_m} \circ \sem{}{s_1} \circ \ldots \circ \sem{}{s_k}\\
        &   & (\override{\ctx}{\{p_1 \mapsto a_1, \ldots, p_n \mapsto a_n\}})
\end{eqnarray*}

For the common special case when the query runs against a single graph, we support
a version in which this graph is mentioned in the query declaration and treated as
the default graph, so it does not need to be explictly mentioned in the patterns.

$$
\sem{}{{\tt create\ query\ }Q\ (ty_1\ p_1, \ldots, ty_n\ p_n)\ {\tt for\ graph}\ G\{ \mathit{body} \}}(\ctx,a_1,\ldots,a_n) = 
$$
$$
\sem{}{{\tt create\ query\ }Q\ (ty_1\ p_1, \ldots, ty_n\ p_n)\ \{ \mathit{body} \}}(\override{\ctx}{\{ DG \mapsto \ctx(G) \}},a_1,\ldots,a_n).
$$

\end{document}